\documentclass[11pt]{article}
\usepackage[totalwidth=385pt,totalheight=580pt]{geometry}
\usepackage{amsmath,amssymb,epsf,graphics,graphicx}
\usepackage{psfrag}
\newcommand{\fr}[2]{{\textstyle\frac{#1}{#2}}}
\newcommand{\ri}{\right}
\newcommand{\lf}{\left}

\newcommand{\rl}{\rule[-0.5cm]{0cm}{1.cm}}
\newcommand{\rll}{\rule[-0.5cm]{0cm}{1.2cm}}

\newcommand\blank[1]{}

\newcommand{\fract}[2]{{\textstyle\frac{#1}{#2}}}

\renewcommand{\hat}{\widehat}
\newcommand\eq{\begin{equation}}
\newcommand\en{\end{equation}}
\newcommand\bea{\begin{eqnarray}}
\newcommand\eea{\end{eqnarray}}
\newcommand\nn{\nonumber}
\newcommand\ba{\(\begin{array}}
\newcommand\ea{\end{array}\)}
\newcommand{\resection}[1]{\setcounter{equation}{0}\section{#1}}

\newcommand{\Z}{{\mathbb Z}}

\newcommand\te{\theta}

\newenvironment{tab}{\linespread{1.0} \begin{table}}{\end{table}%
\linespread{1.3}}
\begin{document}
\begin{titlepage}
\vskip 0.5cm
\begin{flushright}
DCPT-06/41 \\
{\tt hep-th/0612298}
\end{flushright}
\vskip .7cm
\begin{center}
{\Large{\bf Pseudo-differential equations, and the Bethe Ansatz for
    the classical
Lie algebras
}}
\end{center}
\vskip 0.8cm \centerline{Patrick Dorey$^1$, Clare Dunning$^2$,
Davide Masoero$^3$, Junji Suzuki$^4$ and Roberto Tateo$^5$} \vskip
0.9cm \centerline{${}^1$\sl\small Dept.\ of Mathematical Sciences,
University of Durham,} \centerline{\sl\small  Durham DH1 3LE, United
Kingdom\,}
\vskip 0.3cm \centerline{${}^{2}$\sl\small IMSAS, University of
Kent, Canterbury, UK CT2 7NF, United Kingdom}
\vskip 0.3cm \centerline{${}^{3}$\sl\small SISSA, via Beirut 2-4,
34014 Trieste, Italy}
\vskip 0.3cm \centerline{${}^{4}$\sl\small Department of Physics,
Shizuoka University, Ohya 836, SURUGA, Shizuoka, Japan.}
\vskip 0.3cm \centerline{${}^{5}$\sl\small Dip.\ di Fisica Teorica
and INFN, Universit\`a di Torino,} \centerline{\sl\small Via P.\
Giuria 1, 10125 Torino, Italy}
\vskip 0.2cm \centerline{E-mails:}
\centerline{p.e.dorey@durham.ac.uk, t.c.dunning@kent.ac.uk,}
\centerline{ masoero@sissa.it, sjsuzuk@ipc.shizuoka.ac.jp,
tateo@to.infn.it}

\vskip 1.25cm
\begin{abstract}
\noindent
The correspondence between ordinary differential equations and Bethe
ansatz equations for integrable lattice models in their continuum
limits is  generalised to vertex models  related to classical simple
Lie algebras.  New families of  pseudo-differential equations are
proposed, and a link between specific generalised eigenvalue
problems for these equations
and the Bethe ansatz  is deduced. The pseudo-differential operators
resemble in form the Miura-transformed Lax operators studied in
work on generalised KdV equations, classical W-algebras and, more
recently,  in the context of the geometric Langlands correspondence.
Negative-dimension and boundary-condition dualities are also
observed.
\end{abstract}

{\small

 {\bf PACS:} 03.65.-Ge, 11.15.Tk, 11.25.HF, 11.55.DS.

{\bf Keywords:} conformal field theory, Bethe ansatz,
pseudo-differential equations,  spectral problems. }

\end{titlepage}
\setcounter{footnote}{0}
\def\thefootnote{\fnsymbol{footnote}}
%
\resection{Introduction}
A recent observation \cite{Dorey:1998pt} has established an
unexpected link between two dimensional conformal field theory (CFT)
and the theory of  ordinary differential equations. This rests on a
correspondence between the transfer matrix eigenvalues of certain
integrable models (IMs), in their conformal
limits~\cite{Bazhanov:1994ft,Bazhanov:1996dr}, and the spectral
determinants \cite{Sha,Voros} of ordinary differential equations.

The initial results~\cite{Dorey:1998pt,
Bazhanov:1998wj,Suzuki:1999rj,Dorey:1999uk} connected conformal
field theories with Virasoro central charge $c \le 1$ with
Schr{\"o}dinger problems for one-dimensional anharmonic oscillators.
These conformal field theories are naturally associated to the Lie
algebra
$A_1$, but a generalisation to models related to $A_{n-1}$,
with additional extended W-algebra symmetries, was soon
established
\cite{Dorey:1999pv,Suzuki:1999hu,Dorey:2000ma,Bazhanov:2001xm}.
However, contrary to initial expectations, a simple Lie-algebraic
structure did not emerge immediately, and the extension of the
correspondence to the theories associated with other simple Lie
algebras
$\mathfrak{g}$ has proved surprisingly elusive.

The purpose of this paper is to begin to fill this gap, by
establishing a link between CFTs related to the classical simple Lie
algebras and  spectral problems associated with a set of ordinary
(pseudo-) differential equations.

We shall also prove for $\mathfrak{g}=A_{n-1}$, and conjecture  for
the other simple Lie algebras, the existence of closed systems of
functional equations ($\psi$-systems) among uniquely-defined
solutions
$\psi^{(1)},\psi^{(2)}, \dots, \psi^{(rank(\mathfrak{g}))}$ of a set
of  $rank(\mathfrak{g})$ pseudo-differential operators, with each
pair
$\psi^{(a)}$/(operator)$_{a}$ being naturally associated to a node of the
Dynkin diagram.
These $\psi$-systems are very similar to the
systems of functional relations
introduced  by Mukhin and Varchenko in the framework of  the Bethe
ansatz (BA) method for
$\mathfrak{g}$-XXX quantum spin
chains~\cite{Mukhin:2002fp, Mukhin:2002fp1, Mukhin:2002fp2,
Frenkel:2003op}, and in the context of the so-called Miura-opers
related to the geometric Langlands correspondence (see, for example,
\cite{Frenkel:2005fr,Chervov:2006xk}).
This similarity is
related to the fact that
the homogeneous `differential'
parts of the operators studied here resemble, in form, the
Miura-transformed Lax operators  introduced by Drinfel'd and Sokolov
in their studies  of generalised KdV equations and classical
W-algebras~\cite{Drinfeld:1984qv}.

The rest of the paper is organised as follows. \S\ref{BAe} gathers
together  some known, or easily deduced, properties of the Bethe
equations for
$\mathfrak{g}$-type quantum spin chains in their continuum limits.
Our main results are summarised in \S\ref{main}, while extra details and
numerical support for the specific
$A,D,B$ and $C$ proposals  are given in \S\ref{seca} to
\S\ref{secc} respectively, and \S\ref{conclusions} contains our
general conclusions.
There are two appendices: appendix~\ref{appa} deals with the
semiclassical analysis for
$A_1$-related  ODEs  in the presence of string solutions, and
appendix~\ref{appb} describes a simple algorithm  useful for the
numerical solutions of the differential equations.  The algorithm is
a generalisation of Cheng's method from Regge pole
theory~\cite{cheng:1962}, and relies
on an elegant dual formulation of  the relevant  boundary problems.
\section{The Bethe Ansatz equations  and their string solutions}
\label{BAe}
For any simple Lie algebra $\mathfrak{g}$ of type $A_{n-1}$ to
$G_2$, a set of Bethe ansatz equations (BAEs), depending on a set of
$rank(\mathfrak{g})$ twist parameters $ \gamma{=}\{ \gamma_a \}$,
can be written in a universal form as \cite{Schulz:1983,
Babelon:1982gp, Reshetikhin:1986vd, Reshetikhin:LOMI1984,
Reshetikhin:1987nn, Reshetikhin:1987bz}\footnote{ For finite lattice
models, the explicit diagonalisation of the $A_{n-1}$ cases has been
performed through the algebraic Bethe ansatz
 by Schulz~\cite{Schulz:1983} and also by Babelon,
de Vega and Viallet~\cite{Babelon:1982gp}. For $C_n$ and
$D_n$ models, it has been done by
Reshetikhin~\cite{Reshetikhin:1986vd, Reshetikhin:LOMI1984}.
There is a shortcut to reach the same conclusions via the so-called
analytic Bethe ansatz  of Reshetikhin~\cite{Reshetikhin:1987nn}, and
Wiegmann and Reshetikhin~\cite{Reshetikhin:1987bz}.}

\eq
\prod_{ b=1}^{rank(\mathfrak{g})} \Omega^{B_{ab}\gamma_b}_{\phantom
a} \frac {Q^{(b)}_{B_{ab}}(E^{(a)}_{i},\gamma)}
{Q^{(b)}_{-B_{ab}}(E^{(a)}_{i},\gamma)}= -1\,,\qquad i=0,1,2,\dots~
\label{dall0}
\en
where
\eq
Q^{(a)}_k(E,\gamma)=Q^{(a)}(\Omega^k E,\gamma)~,
\en
and the numbers $E^{(a)}_i$ are the -- in general complex -- zeros
of the functions $Q^{(a)}$:
\eq
Q^{(a)}(E^{(a)}_{i},\gamma)=0~.
\label{zq}
\en

 In (\ref{dall0}) and (\ref{zq}) the indices $a$ and $b$ label
the simple roots of the Lie algebra, the matrix  $B_{ab}$ is defined
by
\eq
B_{ab}= { (\alpha_a, \alpha_b) \over  |\hbox{\rm long
roots}|^2}~,~~~a,b=1,2,\dots,rank(\mathfrak{g})
\label{cab}
\en
and the $\alpha$'s  are the simple roots of $\mathfrak{g}$.
The constant
 $\Omega$ is  a pure phase, parameterised in terms of
a real number $\mu{>}0$ as
\eq
\Omega=\exp \left(i {2\pi \over h^{\vee} \mu} \right)
\en
where $h^{\vee}$ is the dual Coxeter number of $\mathfrak{g}$.
Strictly speaking, the BAE (\ref{dall0}) arise from taking a
suitable continuum or field theory limit of the lattice model BAE, in
the fashion
explained in, for example, \cite{Dorey:2004ta}.

The functions $Q^{(a)}$ appearing in (\ref{dall0}) have a
characteristic asymptotic behaviour at large values of  $-E$
\eq
\ln Q^{(a)}(-E,\gamma) =  m_a {\sin(\frac{\pi}{h^{\vee}}) \over \sin
(\frac{\pi}{h^{\vee}} B_{aa})}(-E)^{\mu}+ \dots~~~.
\label{asQ}
\en

For the $A_{n-1}$, $B_n$, $C_n$ and $D_n$ models the sets  $\{m_a
\}$  are given in table~\ref{tb1}.\footnote{ The  constants $\{ m_a
\}$  are  related  to a particular matrix $K_{ab}$ emerging  from
the analysis of the  Bethe ansatz. For  simply-laced algebras,
$K_{ab}$ is proportional to the Cartan matrix and
$\vec{v}{=}(m_1,m_2,\dots, m_r)$ is  its Perron-Frobenius
eigenvector.} The only free parameter --the overall constant $m$ in
table~\ref{tb1}-- depends on the way the conformal field theory
limit is reached.

\begin{table}[tb]
\begin{center}
\begin{tabular}{|c|c|c|}
\hline \rll \small Model  &\small
$h^{\vee}$ &\small  $m_a$  \\
\hline \hline \rl
 $A_{n-1}$ &  $n$  & $m_a=  2m\sin(a {\pi  \over h^{\vee}} )$,~~$(a=1,\dots,n-1)$  \\
\hline \rl
 $D_n$  &  $2n-2$ & $m_{n-1}=m_{n}=m$,~$m_a=2m \sin(a {\pi  \over h^{\vee}} )$,~~$(a=1,\dots,n-2)$ \\
\hline \rl
 $B_n$   &   $2n-1$ & $m_n=m$,~$m_a=2m \sin(a {\pi  \over h^{\vee}} )$,~~$(a=1,\dots,n-1)$ \\
\hline \rl
 $C_n$   &   $n+1$  & $m_a=  2m\sin(a {\pi  \over 2 h^{\vee}} )$,~~$(a=1,\dots,n)$   \\
\hline
\end{tabular}
\end{center}
\caption{\footnotesize Dual Coxeter numbers and
coefficients
$\{ m_a \}$ for models based on classical simple Lie algebras.}
\label{tb1}
\end{table}

The negative real $E$ axis is also the direction of {\em maximal}\/
growth for $\ln Q^{(a)}(E)$ as $|E|\rightarrow \infty$. {}From
(\ref{asQ}), the Hadamard order of $Q^{(a)}$ is therefore $\mu$ and,
in the so-called `semiclassical regime' $0<\mu<1$, $Q^{(a)}$ can be
written as a convergent infinite product over its zeros as
\eq
Q^{(a)}(E,\gamma)=Q^{(a)}(0,\gamma) \prod_{i=0}^{\infty} \left( 1-
{E \over E^{(a)}_i} \right)~.
\label{fac}
\en
It turns out that the Bethe ansatz roots generally
split into multiplets with approximately equal
modulus
$|E_i^{(a)}|$, and that the ground state
of the quantum spin chain corresponds to a `pure'
configuration of roots, containing only multiplets with a common
dimension
\eq
d_a={K \over B_{aa}}~.
\label{dd}
\en
The integer $K$ in (\ref{dd}) depends on the particular spin chain
under discussion, and corresponds to the degree of
fusion~\cite{Kulish:1981gi,Bazhanov:1989yk, Kuniba:1991bx}. For
$\mathfrak{g}{=}A_1$\,, the spin-$j$ $A_1$ quantum chains,
$K{=}d_1{=}2j$.

It is generally expected \cite{Takahashi1972} that for large values of
$i$ the zeros asymptotically tend to the {\it perfect} string
configurations:
\eq
\arg E_i^{(a)} \sim \left( d_a +1  - 2l \right){B_{aa} \pi \over
h^{\vee} \mu}~,~~~l=1,2,\dots,d_a~. \label{arge}~
\en
Appendix A contains some further discussion of the asymptotic
behaviour of these string solutions.

\section{Summary of the  main results}
\label{main}
This paper is about the correspondence between the Bethe ansatz
equations (\ref{dall0}) for $\mathfrak{g}{=}A_{n-1},B_n,C_n,D_n$
 and spectral problems
associated to   solutions
$\psi(x,E, {\bf g})$ of particular  pseudo-differential equations
with vanishing boundary conditions
\eq
\psi(x,E, {\bf g})= o\left( e^{-{ x^{M+1} \over M+1}}
\right)\quad,\quad (M >K/(h^{\vee}{-}K) ) \label{van}
\en
imposed at large $x$ on the positive real axis.
To specify these
equations we
introduce the $n^{\rm th}$-order differential operator
\cite{Dorey:2000ma}
\eq
D_n({\bf g})=D(g_{n-1}-(n{-}1))\,D(g_{n-2}-(n{-}2))\,\dots\,
D(g_1-1)\,D(g_0)~,
\label{dfactdef}
\en
\eq
D(g)=\left(\frac{d}{dx}-\frac{g}{x}\right)~,
\en
with
\eq
{\bf g} {=}\{g_{n-1}, \dots,g_1, g_0 \}~~~,~~{\bf g^{\dagger}} {=}\{
n-1-g_0, n-1-g_1, \dots, n-1  -g_{n-1} \}~. \label{conj}
\en
We also use the inverse differential operator $(d/dx)^{-1}$,
generally defined through its formal action
\eq
\left( { d \over dx} \right)^{-1} x^{s}= {x^{s+1} \over s+1}
\label{def00}~.
\en
The following properties hold
\eq
\left( { d \over dx} \right) \left( { d \over dx} \right)^{-1}
x^{s}= x^{s}~,~~~\left( { d \over dx} \right)^{-1} \left( { d \over
dx} \right) x^{s}= x^{s}~~~(s \ne 0),
\en
and the integration by parts property
\eq
\left( { d \over dx} \right)^{-1} \left[ f(x) {d \over dx} g(x)
\right]= f(x)g(x)-\left( { d \over dx} \right)^{-1} \left[ g(x) {d
\over dx}f(x) \right]
\label{intbyparts}
\en
is  satisfied, provided in the $x$-expansion of
$f(x)g(x)$ about the origin the constant term is absent. In the
following we shall assume  the validity of (\ref{intbyparts}) by
working implicitly with non-integer  values for the parameters
$g_i$ introduced in (\ref{conj}), and by invoking continuity of the
final results in these parameters.

Finally, we define a basic `potential'
\eq
P_K(E,x)= ( x^{h^{\vee} M/K}-E)^K~.
\label{pk}
\en
\\

With this notation in place,
the following  pseudo-differential  equations are the main concern of
this article: \\

{\bf $A_{n-1}$ (su(n))}:
\eq
\Bigl((-1)^{n}D_{n}({\bf g})-P_K(x,E) \Bigr)\psi(x,E,{\bf g} )=0~,~~
\label{sun0}
\en
with  the constraint $\sum_{i=0}^{n-1}g_i{=}\frac{n(n{-}1)}{2}$.\\
\\

{\bf $D_n$ (so(2n))}:
\eq
\left( D_{n}({\bf g^{\dagger}}) \left( \frac{d}{dx} \right)^{-1}
D_{n}({\bf g}) -\sqrt{P_K (x,E)} \left(\frac{d}{dx} \right)
\sqrt{P_K (x,E)} \right)\psi(x,E,{\bf g})=0
\label{so2n0}
\en\\

{\bf $B_n$ (so(2n+1))}:
\eq
\left( D_{n}({\bf g^{\dagger}}) D_{n}({\bf g}) + \sqrt{P_K (x,E)}
\left(\frac{d}{dx} \right) \sqrt{P_K (x,E)} \right)\psi(x,E, {\bf
g})=0
\label{so2n10}
\en\\

{\bf $C_n$ (sp(2n))}:
\eq
 \left( D_{n}({\bf g^{\dagger}}) \left(\frac{d}{dx} \right)D_{n}({\bf g})   -P_{K }(x,E)  { \left(d
\over dx \right)^{-1}}P_{K} (x,E) \right)\psi(x,E, {\bf g})=0
\label{sp2n0}
\en\\

The correspondence we  propose links the ground-state
$Q^{(1)}$'s of (\ref{dall0}) to  particular  solutions
$\psi$ (\ref{van}) of equations (\ref{sun0}--\ref{sp2n0}).

In order to  clarify this statement we  introduce an alternative
basis of solutions  $\{ \chi_i (x, E, {\bf g}) \}$ to
(\ref{sun0}--\ref{sp2n0}), characterised by their behaviour near the
origin
\eq
\chi_i(x, E, {\bf g}) \sim  x^{\lambda_i}+ \dots\quad,\quad x\to 0~,
\label{chib}
\en
where the $\lambda$'s are the ordered $(\lambda_0 <\lambda_1<\dots)$
roots of the appropriate indicial equation (see table~\ref{table2}).
\begin{table}[tb]
\begin{center}
\begin{tabular}{|c|c|}
\hline \rll \small Model  & \small indicial equation  \\
\hline \hline \rl
 $A_{n-1}$ &  $\prod_{i=0}^{n-1} (\lambda -g_i)=0$     \\
\hline \rl
 $D_n$  &  $(\lambda-h^{\vee}/2)^{-1} \prod_{i=0}^{n-1} (\lambda -g_i)(\lambda -h^{\vee}+g_i) =0$ \\
\hline \rl
 $B_n$   &  $\prod_{i=0}^{n-1} (\lambda -g_i)(\lambda -h^{\vee}+g_i) =0$   \\
\hline \rl
 $C_n$   &  $
(\lambda-n)\;\prod_{i=0}^{n-1}(\lambda-g_i)(\lambda-2n+g_i)=0$  \\
\hline
\end{tabular}
\end{center}
\caption{\footnotesize Indicial equations. \label{table2}}
\end{table}

Writing $\psi$ as a linear combination of the $\chi$'s, we have in
general
\eq
\psi(x,E,{\bf g}) =  Q^{(1)}_{[0]}(E, {\bf g}) \, \chi_0 (x, E, {\bf
g})+ Q^{(1)}_{[1]}(E, {\bf g}) \, \chi_1 (x, E, {\bf g})+\dots~.
\label{psichi}
\en
If the  zeros of  $Q^{(1)}_{[0]}(E,{\bf g})$ are $ \{ E_i^{(1)} \}$,
then for $E{\in} \{ E_i^{(1)} \}$ the function $x^{-\lambda_0}
\psi(x,E,{\bf g})$ vanishes exceptionally both at $x{=}\infty$ and
at $x{=}0$. This allows the coefficient $Q^{(1)}_{[0]}(E,{\bf g})$
to be identified with the  spectral determinant for a  boundary
problem defined on the positive real axis (see, for example,
\cite{Sha, Voros}). An alternative (dual) definition of the spectral
functions $Q^{(1)}_{[0]}(E,{\bf g})$ in terms of the adjoint
equations to~(\ref{sun0})--(\ref{sp2n0}) is briefly discussed in
appendix~\ref{appb}.

We claim  that for classical Lie algebras  with
arbitrary degree of fusion
$K$, the ground-state
$Q^{(1)}(E,\gamma)$'s in  (\ref{dall0}) and the functions
$Q^{(1)}_{[0]}(E,{\bf g})$ in (\ref{psichi}) coincide  up to a trivial
normalisation, so that
\eq
{ Q^{(1)}(E, {\bf \gamma}) \over  Q^{(1)}(0, {\bf \gamma}) } = {
Q^{(1)}_{[0]}( E, {\bf g}) \over  Q^{(1)}_{[0]}(0, {\bf g})}~.
\label{eqqq}
\en

{}Moreover, from the WKB approximation
\eq
\ln Q^{(1)}_{[0]}(-E,{\bf g}) = \kappa \;
(-E)^{\hat{\mu}}+\dots~~~~(E\gg 0)
\label{asqq}
\en
with
\eq
\hat{\mu}={ K(M+1) \over h^{\vee} M}~,~~~\kappa=\kappa\left({
h^{\vee} M \over K} , {h^{\vee} \over K} \right)~~
\en
and
\eq
\kappa(a,b)= \int_{0}^{\infty} dx \left( (x^{a}+1)^{b} -x^{ab}
\right)= {\Gamma(1+1/a)\Gamma(1+1/b) \sin(\pi/b) \over
\Gamma(1+1/a+1/b) \sin(\pi/b+\pi/a)}~.
\en
Therefore, in order to have a match between (\ref{asqq}) and (\ref{asQ}), we
must set
\eq
\mu=\hat{\mu}~,~~~m_1=\kappa {\sin(\frac{\pi}{h^{\vee}}B_{11}) \over
\sin (\frac{\pi}{h^{\vee}})}~,~~
\Omega=\exp \left(i {2\pi M\over K(M+1)} \right)~.
\label{Omega}
\en
Given a  particular ordering convention, the relationship
between the twist parameters
$\{ \gamma_a \}$ and the constants $\{ g_a\}$ is given in
table~\ref{table3}.

\begin{table}[tb]
\begin{center}
\begin{tabular}{|c|c|c|}
\hline \rll \small Model  & ordering~~$\forall~i<j$ & \small
$\{ g_a \} \leftrightarrow \{\gamma_a \}$   \\
\hline \hline \rl
 $A_{n-1}$ & $g_i<g_j$ &$\gamma_a= \alpha \left(\sum_{i=0}^{a-1} g_i -
{a(h^{\vee}-1) \over 2} \right)~$     \\
\hline \rl {}  & {} &$\gamma_a= \alpha \left( \sum_{i=0}^{a-1} g_i -
{a \over 2}
h^{\vee} \right)~,~~(a=1,\dots, n-2)$ \\
 $D_n$   & $g_i<g_j<h^{\vee}/2$  & $\gamma_{n-1}= {\alpha \over 2}  \left( \sum_{i=0}^{n-1} g_i - {n
\over 2}  h^{\vee}   \right)$  \\
{} & {} &$\gamma_{n}= {\alpha \over 2} \left( \sum_{i=0}^{n-2} g_i -
   g_{n-1} - {n-2 \over 2}  h^{\vee}   \right)$ \\
\hline \rl
 $B_n$   & $g_i<g_j<h^{\vee}/2$ & $\gamma_a= \alpha \left( \sum_{i=0}^{a-1} g_i - {a
\over 2}
h^{\vee} \right)$   \\
\hline \rl
 $C_n$   & $g_i<g_j<n$& $\gamma_a= \alpha \left(\sum_{i=0}^{a-1} g_i - a n \right)$ \\
{}  & {}&$\gamma_n= { \alpha \over 2} \left( \sum_{i=0}^{n-1} g_i -
n^2 \right)$
 \\
\hline
\end{tabular}
\end{center}
\caption{\label{table3} \footnotesize The relationship between the
set of parameters $\{ g_a \} \leftrightarrow \{\gamma_a \}$ with
$\alpha {=} 2K/Mh^\vee$.}
\end{table}
Various consistency checks, including the
WKB approach and numerical work, support the correspondence both
qualitatively and quantitatively.

Finally, starting from equations (\ref{sun0}) to (\ref{sp2n0}), the
Bethe ansatz equations and table~\ref{table3} were obtained with the
help of a
system of functional relations involving
$\psi^{(1)}(x,E,{\bf g})=\psi(x,E,{\bf g})$ together with other
auxiliary functions $\psi^{(a)}(x,E,{\bf g})$\,,
$(a=2,\dots,rank(\mathfrak{g}))$ (see
\S\ref{seca}--\S\ref{secc} and  \cite{Dorey:1999pv, Dorey:2000ma}).
We set\footnote{In the  $A_{n-1}$ models
$\psi_k^{(1)}(x,E, {\bf g})=\omega^{(n-1)k/2} y_{-k}(x,E, {\bf g})$,
where $y_{k}$ is the function defined in \S3 of \cite{Dorey:2000ma}.
}
\eq
\psi^{(a)}_k= \psi^{(a)}(\omega^{k}x, \Omega^{k}E, {\bf g})
\label{rotated}
\en
where
\eq
\Omega=\exp \left(i {2\pi M\over K(M+1)} \right)
\en
as in (\ref{Omega}), and
\eq
\omega=\Omega^{K/h^{\vee}}=
\exp \left(i {2\pi \over h^{\vee} (M+1)} \right)~.
\en
For the simply-laced algebras the $\psi-$systems can then
be written in
the compact form
\eq
W[\psi_{- \frac{1}{ 2}}^{(a)}, \psi_{\frac{1}{2}}^{(a)}]=
\prod_{b=1}^{rank(\mathfrak{g})} (\psi^{(b)})^{A_{ab}}~,
\label{ssiade}
\en
where $A_{ab}{=} 2\delta_{ab}-2 B_{ab}$ is the incidence matrix of
the corresponding Dynkin diagram and $W$ the Wronskian:
\eq
W[f,g]= f(x) \frac{d}{dx} g(x)- g(x) \frac{d}{dx}f(x)~.
\label{2wr}
\en
Equation (\ref{ssiade}) is proven in \S\ref{anpsi}  for
$\mathfrak{g}{=}A_{n-1}$, and our numerical results indirectly
support the validity of (\ref{ssiade}) for $\mathfrak{g}{=}D_n$.
Currently, we have no analogous pseudo-differential equations for
the exceptional Lie algebras but the similarity between
(\ref{ssiade}), the relations proposed in ~\cite{Mukhin:2002fp,
Mukhin:2002fp1, Mukhin:2002fp2} and the other functional equations
(Y-systems and T-systems) discovered in the framework of integrable
models \cite{Zamolodchikov:1991et, Kuniba:1992ev, Ravanini:1992fi,
Kuniba:1993cn, Kuniba:1993nr,Kuniba:1994na} suggests the validity of
(\ref{ssiade}) in its more general form. For the non simply-laced
algebras our conjectures are

\bea
B_n&:&~~W[
\psi_{-\frac{1}{2}}^{(a)},\psi_{\frac{1}{2}}^{(a)}]=\psi^{(a-1)}
\psi^{(a+1)};~~~ a=1,\dots,n-1,\nn \\
&~&~~ W[ \psi_{-\frac{1}{4}}^{(n)}, \psi_{\frac{1}{4}}^{(n)}]=
\psi^{(n-1)}_{-\frac{1}{4}} \psi^{(n-1)}_{\frac{1}{4}}~.
\eea
(Our  root convention  is $(\alpha_i|\alpha_i){=}2$ for
$i=1,2,\dots,n-1$ and
$(\alpha_n|\alpha_n){=}1$.)\\[-5pt]

\bea
{}~C_n&:&~~W[\psi^{(a)}_{-\frac{1}{4}}, \psi^{(a)}_{\frac{1}{4}}] =
\psi^{(a-1)} \psi^{(a+1)}~;~~~~ a=1,\dots,  n-2, \nn \\
&~&~~W[\psi^{(n-1)}_{-\frac{1}{4}}, \psi^{(n-1)}_{\frac{1}{4}}] =
\psi^{(n-2)}\; \psi^{(n)}_{-\frac{1}{4}}  \psi^{(n)}_{\frac{1}{4}}~,
\nn
\\
&~&~~W[\psi^{(n)}_{-\frac{1}{2}}, \psi^{(n)}_{\frac{1}{2}}] =
\psi^{(n-1)}~.
\eea
(Here, $(\alpha_i|\alpha_i){=}1$ for $i=1,2,\dots,n-1$ and
$(\alpha_n|\alpha_n){=}2$.)\\[-5pt]

\bea
F_4&:&~~W[
\psi_{-\frac{1}{4}}^{(1)},\psi_{\frac{1}{4}}^{(1)}]=\psi^{(2)}~,\nn \\
&~&~~ W[ \psi_{-\frac{1}{4}}^{(2)}, \psi_{\frac{1}{4}}^{(2)}]=
\psi^{(1)}\;\psi^{(3)}_{-\frac{1}{4}} \psi^{(3)}_{\frac{1}{4}}, \nn \\
&~&~~ W[ \psi_{-\frac{1}{2}}^{(3)}, \psi_{\frac{1}{2}}^{(3)}]=
\psi^{(2)}\;\psi^{(4)}, \nn
\\ &~&~~ W[ \psi_{-\frac{1}{2}}^{(4)}, \psi_{\frac{1}{2}}^{(4)}]=
\psi^{(3)}.
\eea
\nobreak (Here, $(\alpha_1,\alpha_1){=}(\alpha_2,\alpha_2){=}1$ and
$(\alpha_3,\alpha_3){=}(\alpha_4,\alpha_4){=}2$.)\\[-5pt]

\bea
G_2&:&~~W[
\psi_{-\frac{1}{2}}^{(1)},\psi_{\frac{1}{2}}^{(1)}]=\psi^{(2)}~,\nn \\
&~&~~ W[ \psi_{-\frac{1}{6}}^{(2)}, \psi_{\frac{1}{6}}^{(2)}]=
\psi^{(1)}\;\psi^{(1)}_{-\frac{2}{6}} \psi^{(1)}_{\frac{2}{6}}.
\eea
(Here, $(\alpha_1|\alpha_1){=}3$ and $(\alpha_2|\alpha_2){=}1$.)\\
\goodbreak

Again, these relations are not proven but we  have indirect
numerical evidence for $\mathfrak{g}{=}B_n$, $C_n$. Further details
and numerical support for the above conjectures are provided in the
following sections, which examine the $A$, $D$, $B$ and $C$ cases in
turn.

\resection{The $A_{n-1}$ models}
\label{seca}
The ODE for the
$A_{n-1}$ models is
\eq
\Bigl((-1)^{n+1}D_n({\bf g})+P_K(x,E) \Bigr)\psi(x,E,{\bf g} )=0~,~~
\label{gnde}
\en
where the operator  $D_n({\bf g})$ and the generalised potential
$P_K(x,E)$ were defined   in \S\ref{main}, and the
additional constraint
\eq
\sum_{i=0}^{n-1}g_i=\frac{n(n{-}1)}{2}
\label{gvan}
\en
ensures that the term $x^{-1}\frac{d^{n-1}}{dx^{n-1}}\,$ is absent.

The function $\psi(x,E, {\bf g})$ is defined to be the most
subdominant solution on the
positive real axis, with asymptotic behaviour, for
$M >K/(h^{\vee}{-}K)$, given by
\eq
\psi(x, E, {\bf g}) \sim {\cal N} \; x^{(1{-}n)M/2}
\exp(-x^{M+1}/(M{+}1)) \label{asy}
\en
as $x \to\infty$. The coefficient
${\cal N}$ represents
an $E$-- and  ${\bf g}$--independent normalisation constant.

The $K{=}1$ cases have been extensively discussed in
\cite{Dorey:1999pv,Suzuki:1999hu,Dorey:2000ma,Bazhanov:2001xm}; they
are related to the $WA_{n-1}$ conformal field theories with integer
Virasoro central charge $c = n{-}1$. Alternatively, at particular
values of the parameters ${\bf g}$ and $M$ they can also be put in
correspondence with the minimal coset conformal field theories
\eq
{(\hat{A}_{n-1})_1 \times (\hat{A}_{n-1})_L  \over
(\hat{A}_{n-1})_{L+1}}~,
\label{coset}
\en
with a simple relationship between $L$ in (\ref{coset}) and the
parameter $M$ in~(\ref{pk}).

The generalisation to integer $K{>}1$ comes from an observation by
Sergei Lukyanov \cite{Luk-private} for the $A_1$ case, for which
numerical and analytic  support was later  provided in
\cite{Dorey:2003sv} and in \cite{Lukyanov:2006gv}. It is reasonable
to  conjecture that this generalisation works both for $A_{n-1}$
with $n{>}2$ and, up to  minor   modifications, for the other models
to be discussed in this paper. In analogy to (\ref{coset}), at
particular values of ${\bf g}$ and $M$, the integer $K{>}1$  cases
should correspond to the cosets
\eq
{\hat{\mathfrak{g}}_K \times \hat{\mathfrak{g}}_L  \over
\hat{\mathfrak{g}}_{K+L}}~,
\label{cosetm}
\en
which describe conformal field theories  with
$\mathfrak{g}{=}A_{n-1},B_n,C_n, D_n$. In appendix~\ref{appa} we
shall explain, in the simplest case, why potentials of such forms
naturally lead to string patterns of roots of the sort mentioned in
the introduction.
\subsection{Negative-dimension dualities}
\label{negative}
It is interesting to  note that there are formal duality relations
among our pseudo-differential equations  involving  negative values
of $n$ and $K$. Consider the $A_{n-1}$ ODEs with the twists ${\bf
g}{=}\{0,1,\dots,n-1\}$:
\eq
\Bigl((-1)^{n+1} {d^{n} \over dx^n}+P_K(x,E) \Bigr)\psi(x,E)=0~.
\label{gnde1}
\en
Setting
$\tilde{\psi}(x,E)= P_K(x,E) \psi(x,E )$ and multiplying from the
left by  $({d \over dx})^{-n}$, the result is
\eq
\Bigl((-1)^{n+1} \left({d \over dx} \right)^{-n} + P_{-K}(x,E)
\Bigr)\tilde{\psi}(x,E )=0~.~~
\label{gnde2}
\en
Comparing (\ref{gnde2}) with (\ref{gnde1}) and taking into account
the boundary conditions, we see that there is a formal duality and
 a spectral equivalence between the initial
$n^{\rm th}$-order ODE and the pseudo-differential equation
(\ref{gnde2}):
\eq
\{ n, M, K \} \leftrightarrow  \{-n,M,-K \}~.
\label{dualan}
\en
Though the above manipulation might look purely formal, it strongly
resembles previously-observed  W-algebra
dualities~\cite{Hornfeck:1994is}:

\eq {(\hat{A}_{-n})_K \times
 (\hat{A}_{-n})_L \over (\hat{A}_{-n}) _{K+L}}
\sim  {(\hat{A}_n)_{-K} \times (\hat{A}_n)_{-L}
 \over (\hat{A}_n)_{-K -L}}~.
\label{dualaa}
\en
A  discussion of the  precise  relation  between $L$
and   the ODE  parameters
$\{ n, M, K, {\bf g} \} $ is not
important for the current naive considerations
and we shall postpone it to the future.

Whilst the duality (\ref{dualan})  remains at the moment a purely
formal observation, the  second duality discussed
in~\cite{Hornfeck:1994is}
\eq
{(\hat{D}_{-n})_K \times (\hat{D}_{-n})_L \over
 (\hat{D}_{-n})_{K+L}} \sim { (\hat{C}_{n})_{-K/2} \times
  (\hat{C}_{n})_{-L/2}
 \over  (\hat{C}_{n})_{-K/2 -L/2}}\,,
\label{dualdc}
\en
will lead us  to  the proposal (\ref{sp2n0})  for the
$C_n$-related equations. (For the  quantisation of the
classical $D_n$ W-algebras in relation to the Miura-opers see, for
example, \cite{Lukyanov:1989gg}.)

 Negative-dimension
dualities resembling those described here are also well-known to
group theorists; many more details can be found in
\cite{CvitanovicWEB}.

\subsection{Auxiliary functions and the $\psi$-system}
\label{anpsi}
The solution $\psi(x,E,{\bf g})$ of (\ref{gnde})  with the
asymptotic behavior (\ref{asy}) is not the only function associated
to the $A_{n-1}$ Bethe ansatz equations. In order to derive the
Bethe ansatz itself a total of $n{-}1$ functions
$\psi^{(1)},\dots,\psi^{(n-1)}$, one for each node of the $A_{n-1}$
Dynkin diagram, should be introduced. These auxiliary functions are
solutions of generally  more-complicated ordinary differential
equations.

Following~\cite{Dorey:2000ma}, all the functions
$\psi^{(a)}(x,E,{\bf g})$ decaying  at large $x$ on  the positive real
axis can be constructed from  $  \psi \equiv \psi^{(1)} $ as
\eq
\psi^{(a)} =
W^{(a)}[\fract{1-a}{2},\fract{3-a}{2},\dots,\fract{a-1}{2} ]
\equiv W^{(a)}[\psi_{{1-a \over 2}},\psi_{{3-a \over
2}},\dots,\psi_{a-1 \over 2} ]~,
\label{psidef}
\en
where $a=1,2,\dots, n{-}1$, $W^{(a)}[f_{1},\dots,f_{a}]$ denotes the
generalised Wronskian of the set of functions $\{f_{a}\}$
\eq
W^{(a)}[f_{1},\dots,f_{a}] = {\bf det} \left[
\left(\vec{f},\frac{d}{dx} \vec{f},\dots,\frac{d^{a-1}
}{dx^{a-1}}\vec{f} \right )\right]
\en
with  $\vec{f}= ( f_1,f_2,\dots, f_{a} )$ (so that $W^{(2)}[f,g] \equiv
W[f,g]$,~cf.\ eq.\,(\ref{2wr})\,),   and $\psi_k$ denotes the `rotated'
solution (\ref{rotated}).

Finally, normalising $\psi^{(1)}$ by
\eq
{\cal N}= { i^{(n-1)/2} \over \sqrt{n}}  ~~\Longrightarrow~~
 \psi^{(n)}(x,E,{\bf g})=1~.
\en

Since $\psi$ is a solution of an $n^{\rm th}$-order ODE, a naive counting of
degrees of freedom shows that the order of the ODE satisfied by
$\psi^{(a)}$ should be
\eq
{ n! \over (n-a)! \;a!}~.
\label{dime}
\en
The $(n-a+1)$ and
the $(a)$ equations are
related by a (${\bf g}
\leftrightarrow {\bf
 g}^{\dagger}$)-conjugation \cite{Dorey:1999pv,Dorey:2000ma}, arising
from the $\Z_2$ symmetry of the Dynkin diagram.
Notice that the  result  (\ref{dime}) exactly matches the dimensions of
the basic representations of $A_{n-1}$; these  are again in
one-to-one correspondence with the nodes of the Dynkin diagram.

Fortunately, in order to derive the Bethe ansatz equations an
explicit knowledge of the remaining $n{-}2$ ODEs is unnecessary: the
derivation of \cite{Dorey:2000ma}  was instead based on the Stokes
relation associated to (\ref{gnde})
\eq
\sum_{k=0}^{n} (-1)^k C^{(k)}(E,{\bf g})\, y_k(x,E,{\bf g}) = 0 ~,
\label{nstokes}
\en
where, according to~\cite{Dorey:2000ma},
\eq
y_k(x,E,{\bf g})=\omega^{(n-1)k/2} \psi(\omega^{-k}x, \Omega^{-k}
E,{\bf g} )~,
\en
 $C^{(0)}(E,{\bf g})=1$ and the Stokes multipliers
$C^{(k)}(E,{\bf g})$ with  $k{>}0$ are analytic functions of $E$ and
${\bf g}$. Stokes relations for the $D_n, B_n$ and $C_n$
equations~(\ref{so2n0}), (\ref{so2n10}) and (\ref{sp2n0})   also
exist, but we have encountered some subtle complications\footnote{In
the $D_n$ case these complications are probably a consequence of the
fact that the ODEs associated to the $\Z_2$-conjugate nodes in the
Dynkin diagram are somehow more  fundamental than eq.~(\ref{so2n0}).
This latter equation  is more naturally associated to the first node
on the `tail' of the diagram.} in generalising the approach of
\cite{Dorey:2000ma} to these cases.

Instead, the strategy  here is   based on the conjectured  validity of a
simple `universal' system of functional equations among the
$\psi^{(a)}$ functions,
 which leads immediately to the Bethe ansatz  equations
and bypasses
the analysis of Stokes relations.

We  shall now  prove the $A_{n-1}$ $\psi$-system~(\ref{ssiade}):
\eq
\psi^{(a-1)} \psi^{(a+1)} = W [\psi_{- \frac{1}{ 2}}^{(a)},
\psi_{\frac{1}{2}
  }^{(a)}]~, \quad  \psi^{(0)}=\psi^{(n)}=1~.
\label{starstar}
\en
The proof is based on the observation that determinants satisfy
functional equations, in particular the so-called Jacobi
identity
\eq
\Delta \; \Delta[p, q|p, q] = \Delta[p|p] \; \Delta[q|q] -
 \Delta[p|q] \; \Delta [q|p]~.
\label{jacobi}
\en
Here,  $\Delta$ is the
determinant of an
$(a+1) \times (a+1)$ matrix and $\Delta[p_1, p_2 |q_1, q_2]$ denotes the
determinant of the same matrix with the $p_{1, 2}-{\rm th }$ rows and
$q_{1,
2}-{\rm th}$ columns removed.

In order to prove (\ref{starstar}) we have to consider  three
different cases: $a{=} 1$, $1 {<} a {<} n {-} 1$ and $a {=} n {-}
1$. The $a{=}1$ case follows from the definition of $\psi^{(2)}$
given in~(\ref{psidef}). Equation~(\ref{starstar}) for $1 {<} a {<}
n {-} 1$ follows from the following chain of identities
\bea
& &\prod_b (\psi^{(b)})^{A_{ab}} =
 W^{(a+1)} [-\fract{a}{2}, -\fract{a-2}{2},\dots, \fract{a-2}{2}, \fract{a}{2}] \; W^{(a-1)} [- \fract{a-2}{2}, \dots, \fract{a-2}{2}]\nn \\
& &\phantom{ccc} =
 (- 1)^{(a - 1)} W^{(a + 1)} [-\fract{a-2}{2} , \dots, \fract{a-2}{2}, - \fract{a}{2}, \fract{a}{2}] \;
 W^{(a - 1)} [-\fract{a-2}{2}, \dots, \fract{a-2}{2} ] \nn \\
& &\phantom{ccc}= (- 1)^{(a-1)} \Delta \, \Delta [a, a + 1|a, a + 1]
\eea
where we have identified
\eq
\Delta \equiv W^{(a + 1)} [-\fract{a-2}{2}, \dots, \fract{a-2}{2}, -
\fract{a}{2}, \fract{a}{2}]
\en
and
\eq
\Delta[a, a + 1|a,a+ 1]=W^{(a - 1)} [- \fract{a-2}{2} , \dots,
\fract{a-2}{2} ].
\en
This is nothing but the LHS of the Jacobi identity
(\ref{jacobi}). Then
an application of the Jacobi identity naturally proves
(\ref{starstar}) in the following way:
\bea
& &\prod_b (\psi^{(b)})^{A_{ab}}
   =  (- 1)^{(a-1)} (\Delta [a|a] \; \Delta [a+1|a+1] -
           \Delta [a|a+1]\; \Delta[a+1|a]) \nn \\
&  &\phantom{ccc} = (- 1)^{(a-1)} {\big(}W^{' (a)} [-
\fract{a-2}{2}, \dots, \fract{a-2}{2} , \fract{a}{2}]\; W^{(a)} [-
\fract{a-2}{2}
  , \dots, \fract{a-2}{2}, - \fract{a}{2}] \nn \\
& &\phantom{cccccccccc}
 - W^{' (a)} [-\fract{a-2}{2}, \dots, \fract{a-2}{2}, -\fract{a}{2}]\; W^{(a)} [-\fract{a-2}{2}, \dots, \fract{a-2}{2}, \fract{a}{2}]{\big  )} \nn \\
& & \phantom{ccc}  =
  W^{' (a)} [-\fract{a-2}{2}, \dots, \fract{a-2}{2}
  , \fract{a}{2}]\; W^{(a)} [-\fract{a}{2}, - \fract{a-2}{2}, \dots,\fract{a-2}{2}] \nn \\
& & \phantom{cccccccccc} - W^{' (a)} [- \fract{a}{2}, -
\fract{a-2}{2},\dots, \fract{a-2}{2} ]\;
W^{(a)} [-\fract{a-2}{2} , \dots, \fract{a-2}{2}, \fract{a}{2}] \nn \\
& &\phantom{ccc} = \psi_{-\frac{1}{2}}^{(a)} \psi_{\frac{1}{2}}^{'
(a)} - \psi_{- \frac{1}{2}}^{' (a)} \psi_{\frac{1}{2}}^{(a)}
  =  W[\psi_{- \frac{1}{2}}^{(a)}, \psi_{\frac{1}{2}}^{(a)}]~.
\eea

Finally, for  $a {=} n {-} 1$ the previous calculation shows that
\eq
\psi^{(n - 2)} \psi^{(n)} = W[\psi_{- \frac{1}{2}}^{(n - 1)},
\psi_{\frac{1}{2}}^{(n - 1)}]~.
\en
Choosing ${\cal N}={ i^{(n-1)/2} \over \sqrt{n}}$
 gives
$\psi^{(n)}{=}1$  and (\ref{starstar}) is proved.

\subsection{The $A_{n-1}$ Bethe ansatz equations}
\label{anBAe}
In this section we shall show that the  BAEs are  a simple consequence
of the $\psi$-system.
First, recall the alternative $\chi$-basis of
solutions~(\ref{chib}) and
the formal ordering  of table~\ref{table3}
\eq
g_i<g_j~~~~\forall \  i<j~.
\label{gordering}
\en
These solutions are defined by their behaviour as $x \to 0$
\eq
\chi_i(x, E, {\bf g}) \sim  x^{\lambda_i}+ O (x^{\lambda_i +
n})~~~,~~~~ \lambda_i=g_i~~.
\en
Next, expand
$\psi(x, E, {\bf g})$ in the $\chi$-basis
\eq
\psi(x, E, {\bf g}) = \sum_{i = 0}^{n - 1} Q_{[i]}^{(1)} (E, {\bf
g}) \chi_i (x, E, {\bf g})~,
\en
and use the property
\eq
\chi_i (\omega^{k} x, \Omega^{k} E, {\bf g}) = \omega^{k \lambda_i}
\chi_i (x, E, {\bf g})~,
\en
to obtain
\eq
\psi_k (x, E, {\bf g}) =  \sum_{i = 0}^{n - 1} Q_{[i]}^{(1)}
   (\Omega^{k} E, {\bf g} )\, \omega^{k \lambda_i} \chi_i (x, E, {\bf g})~.
\en
Now  expanding the determinants in the determinants for this new
basis leads to
\bea
 \psi^{(a)} (x, E, {\bf g}) & = & \sum_{{\bf i}}  \left( \prod_j
   Q_{[i_j]}^{(1)} (\Omega^{j} E, {\bf g}) \, \omega^{j \lambda_{i_j} }
 \right)
    W^{(a)} [\chi_{i_\frac{1-a}{2}}, \dots, \chi_{i_\frac{a-1}{2}}] \nn \\
 & = & {\sum_{\bf i}}'
   Q_{[i_{\frac{1-a}{2}},\dots, i_{\frac{a-1}{2}}]}^{(a)} (E, {\bf g})
   W^{(a)}[\chi_{i_\frac{1-a}{2}}, \dots, \chi_{i_\frac{a-1}{2}}]
\label{expanding}
\eea
where $j=\frac{1-a}{2},  \frac{3-a}{2}, \dots,\frac{a-1}{2}$,
$\sum_{{\bf i}}$ denotes the sum from $0$ to $n{-}1$ of the set $\{
i_j \}$ while in ${\sum_{{\bf i}}}'$ there is the additional
constraint $ 0 \le i_{\frac{1-a}{2}} \le i_{\frac{3-a}{2}} \le \dots
\le i_{\frac{a-1}{2}}$.

A family of $x$-independent equations is obtained by identifying
from the LHS and RHS of (\ref{starstar}) the terms corresponding to
the same power. It is possible a-priori to identify from every
determinant only the highest, second highest, lowest and second
lowest orders. We shall extract the leading orders, though similar
results can be obtained from the subdominant ones.

Setting
\eq
Q_k^{(a)} (E,{\bf g}) = Q^{(a)}_{[0, \dots, a - 1]} (\Omega^{k}
E,{\bf g})~,~~~ \bar{Q}_k^{(a)} (E,{\bf g}) = Q_{[0, \dots, a - 2,
a]}^{(a)} (\Omega^{k} E,{\bf g})
\en
we have
\bea
\psi_k^{(a)} (x, E,{\bf g})& = &\omega^{k \alpha_a} Q_k^{(a)}(E,{\bf
g})
W^{(a)} [\chi_0, \dots, \chi_{a- 1}] + \nn \\
& &  \omega^{k (\alpha_{a + 1} -\lambda_{a - 1})} \bar{Q}_k^{(a)}
(E,{\bf g}) W^{(a)} [\chi_0, \dots, \chi_{a - 2}, \chi_a] + \dots
\nonumber~.\\
& & \label{psiexpansion}
\eea
The  orders of the first and the second terms in
(\ref{psiexpansion}) are
 $\alpha_a{-} a (a{-} 1) / 2$ and $\alpha_{a + 1} {-} \lambda_{a - 1} {-} a (a {-} 1) /
2$ respectively, where
\eq
\alpha_a = \sum_{i = 0}^{a - 1} \lambda_i~.
\en
Substituting (\ref{psiexpansion}) into (\ref{starstar}), comparing the
leading terms of  both sides for small $x$ and using the relation
\begin{align*}
W^{(a+1)} W^{(a-1)} &= W[W^{(a)}, \hat{W}^{(a)}]~, \\
W^{(a)}&:=W^{(a)} [\chi_0, \dots, \chi_{a- 1}]~,&
\hat{W}^{(a)}&:=W^{(a)} [\chi_0, \dots, \chi_{a-2},\chi_a]&
\end{align*}
(also proved through the Jacobi identity), we find
\eq
 Q^{(a + 1)} Q^{(a - 1)}  =
   \omega^{\frac{1}{2} (\lambda_{a} -
   \lambda_{a-1})} Q_{- \frac{1}{2}}^{(a)}
   \bar{Q}_{\frac{1}{2}}^{(a)} - \omega^{\frac{1}{2} (\lambda_{a-1}-\lambda_{a} )}
   Q_{\frac{1}{2}}^{(a)}  \bar{Q}_{-
   \frac{1}{2}}^{(a)}~,~~
\en
$Q^{(0)}=1$. Finally, let $E^{(a)}_i$ be a zero of $Q^{(a)}(E, {\bf
g})$. Evaluating the above equation at $E = \Omega^{1 / 2}
E^{(a)}_i$ and at $E = \Omega^{- 1/ 2} E^{(a)}_i$,  and by dividing
the two equations thus obtained, we find the $A_{n-1}$ Bethe Ansatz
equations
\eq
\prod_{ b=1}^{n-1} \Omega^{B_{ab}\gamma_b}_{\phantom a} \frac
{Q^{(b)}_{B_{ab}}(E^{(a)}_{i})} {Q^{(b)}_{-B_{ab}}(E^{(a)}_{i})}=
-1\,,\qquad i=0,1,2,\dots~,
\label{dall1}
\en
where
\eq
\gamma_a={ 2 K \over M h^{\vee}}\left(\sum_{i = 0}^{a - 1}
g_i-a{(n-1) \over 2} \right)~.
\label{gaan}
\en
Notice that in writing relation (\ref{gaan})  we have  used the
identity
 \eq
{ K \over M h^{\vee}}(\lambda_a-\lambda_{a-1})= - \sum_{b=1}^{n-1}
B_{ab} \; \gamma_b~,
\en
the ordering (\ref{gordering}) and imposed  the constraint
$\gamma_0{=}\gamma_n{=}0$.

As  shown in table~\ref{taba4}, for $K{=}1$ there is  very good
agreement between  the IM results obtained from the solution of a
suitable non-linear integral equation~(NLIE)~(see \S6 in
\cite{Dorey:2000ma}) and the  direct numerical solution of the ODE.
\begin{tab}[ht]
\begin{center}
\begin{tabular}{|l | l | l|}
\hline ~Level &~~~~~~~~$A_4$ NLIE &~~~ODE numerics   \\
\hline
~~$E_0^{(1)}$~~&~~~ 14.0495626907~~~ &~~~14.0495626922~~~\\
~~$E_1^{(1)}$~~&~~~ 47.7146839354~~~ &~~~47.7146839363~~~\\
~~$E_2^{(1)}$~~&~~~ 95.1785845453~~~ &~~~95.1785845456~~~\\
~~$E_3^{(1)}$~~&~~~ 154.202021469~~~ &~~~154.202021470~~~\\
~~$E_4^{(1)}$~~&~~~ 223.483044292~~~ &~~~223.483044292~~~\\
\hline \end{tabular} \caption{\footnotesize Comparison of IM results
($A_4$ NLIE) with the direct numerical solution of the $A_4$ ODE
with $K{=}1$, $M{=}10/21$ and ${\bf g}{=} ( 0.2,1.02,2.3,3.421)$.
\label{taba4}}
\end{center}
\end{tab}
Table~\ref{taba42} also shows the good agreement between the IM results
obtained using the  spin-$1$ NLIEs \cite{Suzspin,Dunning:2002tt}
 and the exact solution of the ODE with $K{=}2$ (see \S\ref{b1}\,).

\begin{tab}[ht]
\begin{center}
\begin{tabular}{|l | l | l|}
\hline Level &~~~~~~$A_1$ NLIE with $K{=}2$ &~~~~~~~~~~ODE numerics   \\
\hline ~$E_0^{(1)}$~~& $1.49259741085 \pm  1.60304589242i$~
&$1.49259741085 \pm  1.60304589242 i$~~~\\
~$E_1^{(1)}$~~& $2.31180377628 \pm 2.38537059826i$~
&$2.31180377628 \pm 2.38537059826i$~~~\\
~$E_2^{(1)}$~~& $2.91183770898 \pm  2.97068128676i$~
&$2.91183770898 \pm  2.97068128676i$~~~\\
~$E_3^{(1)}$~~&  $3.40837129214 \pm 3.45880577384i$~
&$3.40837129216\pm 3.45880577388i$~~~\\
~$E_4^{(1)}$~~&  $3.84143464742 \pm 3.88626414305i$~ &$3.84143464640\pm3.88626414641i$~\\
\hline \end{tabular} \caption{\footnotesize Comparison of IM results
($A_1$ NLIE) with the exact solution of the $A_1$ ODE with $K{=}2$,
$M{=}1$ and $g_0 = 0$.  The set $\{2 E_i^{(1)}\}$ are the exact
eigenvalues of the $B_1$ linear ODE of \S\ref{b1}. \label{taba42}}
\end{center}
\end{tab}

\resection{The $D_n$ models}
\label{secd}
The $D_n$ pseudo-differential equation~(\ref{so2n0}) is
\eq
\left( D_{n}({\bf g^{\dagger}}) \left( \frac{d}{dx} \right)^{-1}
D_{n}({\bf g}) - \sqrt{P_{K }(x,E)} \left(\frac{d}{dx} \right)
\sqrt{P_{K }(x,E)} \right)\psi(x,E,{\bf g})=0~.
\label{so2n2}
\en
Following the  $A_{n-1}$ example, we  start from the solution
$\psi(x,E,{\bf g})$ of (\ref{so2n2}) with asymptotic
behaviour
\eq
\psi(x,E,{\bf g}) \sim   {\cal N} \; x^{-h^{\vee} M/2} \exp \left(-
{x^{M+1} \over M+1} \right)\quad,\quad (M >K/(h^{\vee}{-}K) )
\en
as $x \to\infty$ on the positive real axis, and introduce the
alternative basis of solutions~(\ref{chib})
\eq
\chi_i (x, E, {\bf g})~,~~~ i = 0,1, \dots, 2n - 1
\en
characterised by their behaviour near the origin
\eq
\chi_i(x, E, {\bf g}) \sim  x^{\lambda_i}+ O (x^{\lambda_i +
2n-1})~~~,~~~~ \left\{
\begin{array}{ll}
\lambda_i=g_i~, & {\rm for}~ i \le n-1, \\
\lambda_i=h^{\vee}-g_{2n-1-i}~, & {\rm for}~ i>n-1.
\end{array}
\right.
\label{lambdag}
\en

In (\ref{lambdag}) the parameters $\lambda_i$ represent the $2n$
solutions of the indicial equation (see table~\ref{table2}) with the
ordering
\eq
~~~~g_i< g_j\le h^{\vee}/2~,~~~ \lambda_i <
\lambda_j~,~~~~\forall \ i<j~.
\en
\subsection{ The $\psi$-system and the $D_n$ Bethe ansatz equations}
\label{so2npsi}
To extract the $D_n$ BAE, we start from
\eq
\psi^{(1)}_k= \psi_k(x,E, {\bf g})=  \psi(\omega^{k} x,\Omega^{k}
E,{\bf g} )~,
\en
where $\Omega$ and $\omega$ are as defined in~(\ref{Omega}) and
(\ref{rotated}), and assume  the validity, for a suitable value of
the normalisation constant ${\cal N}$, of the $\psi$-system
 (\ref{ssiade})
\bea
W[ \psi^{(a)}_{-\frac{1}{2}} ,\psi^{(a)}_{\frac{1}{2}} ] &=&
\psi^{(a-1)} \psi^{(a+1)}~,~~a=1,\dots,n-3 \nn \\
W[\psi^{(n-2)}_{-\frac{1}{2}} ,\psi^{(n-2)}_{\frac{1}{2}} ] &=&
\psi^{(n-3)} \psi^{(n-1)} \psi^{(n)}, \\
 W[ \psi^{(n-1)}_{-\frac{1}{2}} ,\psi^{(n-1)}_{\frac{1}{2}} ]&=&
 W[ \psi^{(n)}_{-\frac{1}{2}} ,\psi^{(n)}_{\frac{1}{2}} ]=
\psi^{(n-2)}. \nn
\label{dnpsi}
\eea
Equations (\ref{dnpsi}) and  Jacobi identity (\ref{dnpsi}) used in
reverse imply the following relations linking   the  remaining
functions
$\psi^{(a)}(x,E,{\bf g})$ to $\psi^{(1)}(x,E,{\bf g})$:
\eq
\phi^{(a)} \equiv \psi^{(a)}=
W^{(a)}[\psi_{\frac{1-a}{2}},\psi_{\frac{3-a}{2}},\dots,
\psi_{\frac{a-1}{2}}]~,~~~a=1,\dots, n-2~,
\label{df1}
\en
\eq
\phi^{(n-1)} \equiv
\psi^{(n-1)}\psi^{(n)}=W^{(n-1)}[\psi_{\frac{2-n}{2}},\psi_{\frac{4-n}{2}},\dots,
\psi_{\frac{n-2}{2}}]~,
\label{df2}
\en
and
\eq
\phi^{(n)} \equiv   \psi^{(n-1)}_{-\frac{1}{2}}
\psi^{(n-1)}_{\frac{1}{2}}+ \psi^{(n)}_{-\frac{1}{2}}
\psi^{(n)}_{\frac{1}{2}}  =
W^{(n)}[\psi_{\frac{1-n}{2}},\psi_{\frac{3-n}{2}},\dots,
\psi_{\frac{n-1}{2}}]~.
\label{df3}
\en
Now notice that the auxiliary functions $\phi^{(a)}(x,E,{\bf g})$
defined in (\ref{df1}), (\ref{df2}) and (\ref{df3}) satisfy an
$A$-type $\psi$-system
\eq
\phi^{(a-1)} \phi^{(a+1)} =W[ \phi^{(a)}_{-\frac{1}{2}}
,\phi^{(a)}_{\frac{1}{2}} ]~,~~~a=1,\dots, n-1.
\label{antype}
\en
Therefore, the  arguments applied in \S\ref{anBAe} go through in the
same way:
\bea
\!\!\!\phi_k^{(a)} (x, E,{\bf g})& =
 &\omega^{k \alpha_a} \hat{Q}_k^{(a)}(E,{\bf g}) W^{(a)} [\chi_0, \dots, \chi_{a- 1}] + \nn \\
& &  \omega^{k (\alpha_{a + 1} -\lambda_{a - 1})} \bar{Q}_k^{(a)}
(E,{\bf g}) W^{(a)} [\chi_0, \dots, \chi_{a - 2}, \chi_a] + \dots
\label{above1}
\eea
($a=1,2,\dots,n-1$). The  orders of the first and the second terms
in (\ref{above1}) are given  by $\alpha_a{-} a (a {-} 1) / 2$ and
$\alpha_{a + 1} {-} \lambda_{a - 1} {-} a (a {-} 1) / 2$
respectively, with $\alpha_a = \sum_{i = 0}^{a - 1} \lambda_i$ and
\eq
\hat{Q}_k^{(a)} (E,{\bf g}) = \hat{Q}^{(a)}(\Omega^{k} E,{\bf g})~,
~~~ \bar{Q}_k^{(a)} (E,{\bf g})
=\bar{Q}^{(a)} (\Omega^{k} E,{\bf g})~
\en
are  $\phi$-related spectral determinants.

Using equation (\ref{antype}) we establish  the following  identity
among $\phi$-related spectral determinants
\eq
 \hat{Q}^{(a + 1)} (E) \hat{Q}^{(a - 1)} (E) =
   \omega^{\frac{1}{2} (\lambda_{a} - \lambda_{a-1})} \hat{Q}_{- \frac{1}{2}}^{(a)}(E)
   \bar{Q}_{\frac{1}{2}}^{(a)}(E) - \omega^{\frac{1}{2} (\lambda_{a-1}-\lambda_{a} )}
   \hat{Q}_{\frac{1}{2}}^{(a)} (E) \bar{Q}_{- \frac{1}{2}}^{(a)}(E);
\en
which  leads to
\eq
{\hat{Q}^{(a- 1)}_{-\frac{1}{2}} (E^{(a)}_i) \over \hat{Q}^{(a -
1)}_{\frac{1}{2}} (E^{(a)}_i)} {\hat{Q}^{(a)}_1( E^{(a)}_i) \over
\hat{Q}^{(a)}_{-1} ( E^{(a)}_i)}
 {\hat{Q}^{(a+ 1)}_{-\frac{1}{2}} (E^{(a)}_i)
    \over
 \hat{Q}^{(a + 1)}_{\frac{1}{2}}(E^{(a)}_i)}  =-
\Omega^{ {\alpha \over 2} (\lambda_a-\lambda_{a-1})}
\label{fBA}
\en
with $a=1,2,\dots, n-1$ and  $\alpha{=}{2 K \over M h^{\vee}}$. We
then make the following identifications
\bea
\hat{Q}^{(0)}(E,{\bf g}) &=& Q^{(0)}(E,{\bf g})=1;\\
 \hat{Q}^{(a)}(E,{\bf g}) &=&
Q^{(a)}(E,{\bf g})~,~~~(a=1,\dots,n-1) \label{DQ1}; \\
\hat{Q}^{(n-1)}(E,{\bf g}) &=&
Q^{(n-1)} (E,{\bf g})\; Q^{(n)} (E,{\bf g})~~ \label{DQ2}; \\
\hat{Q}^{(n)}(E,{\bf g}) &=& Q^{(n-1)}_{-\frac{1}{2}} (E,{\bf g})\;
Q^{(n-1)}_{\frac{1}{2}} (E,{\bf g}),
\label{DQ3}
\eea
which reflect the relations among the $\phi$'s and the $\psi$'s in
the above. In  (\ref{DQ3}) we have implicitly assumed
\eq
\psi^{(n)}(x,E,{\bf g})= o(\psi^{(n-1)}(x,E,{\bf g})) ~
\en
as $x \rightarrow 0$ and also  that (see the discussion  in
\S\ref{sd2})
\eq
Q^{(n-1)} (E,\{g_0,\dots,g_{n-2},g_{n-1} \}) \equiv Q^{(n)}(
E,\{g_0,\dots,g_{n-2},h^{\vee}-g_{n-1} \})~.
\label{dndd1}
\en
Plugging relations (\ref{DQ1}), (\ref{DQ2}) and (\ref{DQ3})  into
(\ref{fBA}) and using (\ref{dndd1}) is it easy to check that
(\ref{fBA}) can be recast in the universal form (\ref{dall0})
\eq
\prod_{ b=1}^{n} \Omega^{B_{ab}\gamma_b}_{\phantom a} \frac
{Q^{(b)}_{B_{ab}}(E^{(a)}_{i},\gamma)}
{Q^{(b)}_{-B_{ab}}(E^{(a)}_{i},\gamma)}= -1\,,\qquad i=0,1,2,\dots~
\label{dall2}
\en
where $B_{ab}$ is the $D_n$-related  matrix defined according to
(\ref{cab}) and we have imposed the extra condition
\eq
\gamma_{n}-\gamma_{n-1}=n-1-g_{n-1}
\label{gammann}
\en
 to fix the exact $\{ g_a \} \leftrightarrow
\{ \gamma_a \}$ relation as given in  table~\ref{table3}.

Condition (\ref{gammann}) guarantees that when $g_{n-1}{-}n{+}1{=}0$
the operator $(d/dx)^{-1}$ in (\ref{so2n2}) acts directly on a
$d/dx$ and the relevant equation reduces to an $(2n{-}1)$-order ODE.
When this occurs $\gamma_n{=}\gamma_{n-1}$ and so the pair of
$\Z_2$-conjugate nodes of the Dynkin diagram are `twisted' in
exactly  the  same way. Further, a change of   sign in the RHS of
(\ref{gammann}) swaps $\gamma_n$ and $\gamma_{n-1}$, a property that
naturally reflects the presence of the $\Z_2$-symmetry in the $D_n$
Dynkin diagram. All these properties  are  confirmed by the analysis
of \S\ref{sd2}  and \S\ref{a3d3} and by $12$-digits of numerical
agreement  at $K{=}1$ with  ${\bf g}{=} \{ 0,1,\dots,n-1 \}$,
\eq
\gamma_a= \alpha \left({(a-1)a \over 2} - a {h^{\vee} \over 2}
\right)~,~~~ \gamma_{n}=\gamma_{n-1}=- \frac{n h^{\vee}}{4}~,
\en
between NLIE and ODE results.

Table~\ref{tabd4} shows  the (still-excellent) agreement at  $K{=}1$
away from the $\gamma_n=\gamma_{n-1}$ surface. Appropriate $K{>}1$
NLIEs are unknown but numerical results qualitatively reproduce the
expected IM scenario of \S\ref{BAe}. Further analytic support to the
proposed ODE/IM  correspondence  for $D_n$ is given in \S\ref{sd2},
\S\ref{a3d3} and \S\ref{sgso2n} below.

\noindent
\begin{tab}[ht]
\begin{center}
\begin{tabular}{| l | l | l | }
\hline ~Level &~~~~~~~~$D_4$ NLIE &~~~ODE numerics   \\ \hline
~~$E_0^{(1)}$~~&~~~17.8625636061~~~ &~~~17.8625636061~~~\\
~~$E_1^{(1)}$~~&~~~50.2942213433~~~ &~~~50.2942213430~~~\\
~~$E_2^{(1)}$~~&~~~92.8267466445~~~ &~~~92.8267466442~~~\\
~~$E_3^{(1)}$~~&~~~143.348705065~~~ &~~~143.348705065~~~\\
~~$E_4^{(1)}$~~&~~~200.738324171~~~ &~~~200.738324172~~~\\
\hline \end{tabular} \caption{\footnotesize Comparison of IM results
($D_4$ NLIE) with the direct numerical solution of the $D_4$
pseudo-differential equation with  $K{=}1$, $M{=}1/3$ and ${\bf g}
=(0.2, 1.1,2.3,2.95 )$. \label{tabd4}}
 \end{center}
\end{tab}

\subsection{Example 1: $D_2 \sim A_1 \oplus A_1$}
\label{sd2}

The  $D_2$ algebra can be decomposed into  a pair of  independent
$A_1$ algebras, mirroring an analogous factorisation in the BAE. In
this section we shall prove that the solution $\psi(x,E,{\bf g})$ to
(\ref{so2n2}) with $n{=}2$ is   the  product  of two solutions of
$A_1$-related ODEs.

We start from the general $D_2$ equation:
\bea
\left ( \left( {d \over dx} + {g_0 \over x} \right)\left( {d \over
dx} + {g_1-1 \over x} \right)  \left( {d \over dx} \right)^{-1}
\left( {d \over dx} - {g_1-1 \over x} \right)\left( {d \over dx} -
{g_0 \over x} \right) \right.   \nonumber \\
\left.  - \sqrt{P_K(x,E)}  \left({d \over dx}\right)
\sqrt{P_K(x,E)} \right ) \psi(x,E,\mathbf{g})=0~.
\eea
Expanding  and integrating by parts, we obtain an equivalent
equation:
\bea \label{eqn:d2expanded} & &\hspace{-1cm} \left ( -{d^3 \over dx^3}+
4P(x,E,g' ){d \over dx} +  2{dP \over dx}(x,E, g') \right. \nonumber \\
& &\phantom{ccc} \left. +  {
(1-g_0)^2(1-g_1)^2 \over x^2} \left( {d \over dx} \right)^{-1}{1
\over x^2} \right) \psi(x,E, \mathbf{g})=0
\eea
where for this subsection it is convenient to
define $P(x,E, k)= {1 \over 4}[(x^{2M/K} -
E)^K + {k \over x^2}]$\,, and
$g'=g_0^2 - 2 g_0 +g_1^2 -2g_1 +1$.

We now set $ Z(x) = \chi_1(x)\chi_2(x)$, a product of the solutions
of two $A_1$ (spin-${1 \over 2}$) equations, which for general
$\rho$ and $\sigma$ satisfy
\bea
{d^2 \over dx^2} \chi_1(x,E,\rho)&=&
P(x,E, \rho)\,\chi_1(x,E,\rho)~, \label{ab} \\
{d^2 \over dx^2}\chi_2(x,E,\sigma) &=&P(x,E,
\sigma)\,\chi_2(x,E,\sigma)~. \label{ab2}
\eea
To show that $Z(x)$ satisfies  equation (\ref{eqn:d2expanded}), we
differentiate and repeatedly use  the $A_1$ equations to find
\bea
{d^3 Z \over dx^3} &=& 2{d P \over dx}(x,E, {\rho + \sigma \over
2})\,Z + 2P(x,E, {\rho + \sigma
\over 2})\,{dZ \over dx} + 2P(x,E, \rho)\,\chi_1 {d\chi_2 \over dx}\nn\\
 & & \qquad \qquad + 2P(x,E, \sigma)\, {d\chi_1 \over dx} \chi_2~.
\label{eqn:d2third}
\eea
If we now define the Wronskian
\eq
W=\chi_1 {d\chi_2 \over dx} -{d \chi_1 \over dx} \chi_2
\en
and use
\eq
\chi_1 {d\chi_2 \over dx} =  {1 \over 2} \left({dZ \over dx}  + W
\right) \quad ,\quad {d \chi_1 \over dx} \chi_2={ 1 \over 2}
\left({dZ \over dx} - W \right)~,
\en
then (\ref{eqn:d2third}) can be written as
\eq
{d^3 Z \over dx^3}=2{dP \over dx}(x,E, {\rho + \sigma \over 2 } )\,Z
+ 4P(x,E, {\rho + \sigma \over 2 } )\,{dZ \over dx}  + {\rho -
\sigma \over 4 x^2} \,W~.
\en
In order to express $W$ in terms of $Z$ we  differentiate, apply
(\ref{ab}) and (\ref{ab2}), and then integrate:
\eq
{dW \over dx}=\chi_1 {d^2\chi_2 \over dx^2}-{ d^2\chi_1 \over
dx^2}\chi_2={\sigma - \rho \over 4 x^2}Z \quad \rightarrow \quad
W=(\sigma - \rho) \left( {d \over dx} \right)^{-1} {Z \over 4 x^2}~.
\en
 The resulting equation for $Z$,
\begin{align} \label{eqn:a1a1}
& \!\!\!\!\! \!\!\!\!\! \!\!\!\!\! \left( -{d^3 \over dx^3} +
4P(x,E, {\rho + \sigma \over 2})  {d
\over dx}+ 2{dP \over dx}(x,E, {\rho + \sigma \over 2})  \right. \nonumber \\
&\phantom{cccccccccccc} \left. - {(\rho - \sigma)^{2} \over
16x^2}\left( {d \over dx} \right)^{-1} {1 \over x^2} \right) Z(x, E,
\rho,\sigma)=0\,,
\end{align}
exactly matches  equation (\ref{eqn:d2expanded}) provided  the
following relations between $g_0$ and $g_1$, and $\rho$ and $
\sigma$ hold:
\bea
\rho + \sigma &=&2( g_0^2 -2g_0 + g_1^2 -2g_1 +1) \\
\frac{\rho -\sigma}{4}&=& (g_0-1)(g_1 -1)~.
\eea
If $\rho{=} \sigma$ then either $g_0$ or $g_1$ has to be zero and
the integral operator in the $D_2$ equation acts on a total
derivative. This observation agrees with the discussion in
\S\ref{so2npsi} about the relation between $(d/dx)^{-1}$ and an
asymmetric choice  of the twists  $\gamma_n$ and $\gamma_{n-1}$.

\subsection{Example 2: $D_3 \sim A_3$}
\label{a3d3}
The BAE for  $A_3$ and  $D_3$ are the same under identification of
Bethe roots. It is therefore interesting to discuss the exact
correspondence  between  the two models at the level of the
pseudo-differential equations. Actually, it was  the study of  this
case that lead us  to the general $D_n$-related equations.

We start from the observation that the solution $\psi$ of the
$A_3$-related ODE
\eq
\Bigl(D_{4}({\bf g})-P_K(x,E) \Bigr)\psi(x,E,{\bf g} )=0~,~~
\label{su40}
\en
is associated  to the first node of the $A_3$ Dynkin diagram, while
the solution of the $D_3$ equation
\eq
\left( D_{3}({\bf { \bar{g}}^{\dagger}}) \left( \frac{d}{dx}
\right)^{-1} D_{3}({\bf \bar{g}}) - \tau \sqrt{P_K (x,E)}
\left(\frac{d}{dx} \right) \sqrt{P_K (x,E)} \right)\phi(x,E,{\bf
\bar{g}})=0
\label{so6n0}
\en
is more naturally associated to the central node of the $D_3{=}A_3$
Dynkin diagram. (A constant factor $\tau$  has been included in
eq.~(\ref{so6n0})  to  take into account the possibly-different
normalisations for the $E$ parameters.)

Therefore we are looking  for a relationship between $\phi(x,E,{\bf
\bar{g}})$ and
\eq
\psi^{(2)}(x,E,{\bf g})=W[ \psi^{(1)}_{-\frac{1}{2}},
\psi^{(1)}_{\frac{1}{2}}]\,.
\en

For simplicity,  we perform  the calculation at ${\bf
g}{=}\{0,1,2\}$. We set
\eq
\psi^{(2)}(x,E)=[0,1]
\en
where we have introduced  the shorthand notation
\eq
[i,j]= \left(\frac{d^i}{dx^i} \psi^{(1)}_{-\frac{1}{2}} \right)
 \left(\frac{d^j}{dx^j}  \psi^{(1)}_{\frac{1}{2}} \right)-\left(\frac{d^j}{dx^j} \psi^{(1)}_{-\frac{1}{2}} \right)
 \left(\frac{d^i}{dx^i}  \psi^{(1)}_{\frac{1}{2}} \right)
\en
so that
\eq
\frac{d}{dx} [i,j]= [i+1,j]+[i,j+1]\quad,\quad[i,i]=0~,
\en
and
\eq
{ d^4 \over dx^4} \psi^{(1)}_{\pm \frac{1}{2}}(x,E)  =- P_K(x,E)
\psi^{(1)}_{\pm \frac{1}{2}}(x,E)~.
\en
Taking derivatives five times and using the above equation we have
\bea
\psi^{(2)}&=&[0,1] ~,~~ \frac{d}{dx}\psi^{(2)}=[0,2] ~,~~ \frac{d^2}{dx^2}\psi^{(2)}=[1,2]+[0,3] ~, \nn \\
\frac{d^3}{dx^3}\psi^{(2)}&=&2[1,3]+[0,4]=2[1,3]- P_K\;[0,0]=2[1,3] ~, \nn \\
\frac{d^4}{dx^4}\psi^{(2)}&=&2[2,3]+2[1,4]= 2[2,3]+ 2P_K \psi^{(2)} ~, \nn \\
\frac{d^5}{dx^5}\psi^{(2)}&=&2[2,4]+ 2\frac{d}{dx}(P_K \psi^{(2)})
=2P_K \frac{d}{dx}\psi^{(2)} + 2\frac{d}{dx}(P_K \psi^{(2)}) ~,
\eea
and therefore obtain the desired ODE
\eq
\left( \frac{d^5}{dx^5} - 2 \sqrt{P_{K }(x,E)} \frac{d}{dx}
\sqrt{P_{K }(x,E)} \right)\psi^{(2)}(x,E)=0~,
\en
which matches (\ref{so6n0}) at $\tau{=}2$.

 We have also checked that the  solution
$\psi^{(1)}(x,E,{\bf g})$ of the more general $A_3$-related differential
equation~(\ref{su40}) leads to  a function $\psi^{(2)}(x,E,{\bf g})=
W[ \psi^{(1)}_{-\frac{1}{2}}, \psi^{(1)}_{\frac{1}{2}}]$, which is the
solution
of
\eq
\left( D_{3}({\bf {\bar{g}}^{\dagger}}) \left( \frac{d}{dx}
\right)^{-1} D_{3}({\bf \bar{g}}) -2 \sqrt{P_K (x,E)}
\left(\frac{d}{dx} \right) \sqrt{P_K (x,E)}
\right)\psi^{(2)}(x,E,{\bf g})=0~.
\label{so6n0a}
\en

As already seen in \S\ref{sd2},  in order to recast the resulting
equation in the factorised form (\ref{so6n0a}) one has  perform a
number of integrations by parts.

The  exact  relation between the $A_3$ and $D_3$ sets of parameters
is
\bea
2 g_0 &=& 1+\bar{g}_0+\bar{g}_1-\bar{g}_2~,\nn \\
2 g_1 &=& 1+\bar{g}_0-\bar{g_1}+\bar{g}_2~, \\
2 g_2 &=& 1-\bar{g}_0+\bar{g_1}+\bar{g}_2~.\nn
\eea

\subsection{Relationship with the sine-Gordon model}
\label{sgso2n}
The reader may have noticed that the sets of numbers $\{ m_a\}$
for the $D_n$ and $B_n$ models summarised in table~\ref{tb1} match
the mass spectra of the sine-Gordon model at particular values of
the coupling constant. From the sine-Gordon point of view the
$D_n$-related mass spectrum  emerges at the reflectionless  points
where the scattering between the solitons becomes purely diagonal.
This link between the sine-Gordon model, affine Toda field theories
and perturbed coset CFTs has been discussed in various
places~\cite{Klassen:1989ui, Braden:1989bu,
Tateo:1994pb,Dorey:1996ms}.

In this section we would like to point out that there is a simple
connection between equation~(\ref{so2n2}) taken at
\eq
K=1~,~~M=1/(n-1)~,~~{\bf g}=\{0,1,\dots,n-1\}
\label{values}
\en
and the Schr{\"o}dinger  problem  associated, through the first
instance  of the ODE/IM
correspondence~\cite{Dorey:1998pt,Bazhanov:1998wj}, to the CFT limit
of the sine-Gordon model. We start from (\ref{so2n2}) with parameters
(\ref{values}):
\eq
\left(-{d^{2n-1} \over dx^{2n-1}} +(x^2-E) {d \over dx} + x \right)
\chi(x,E)=0
\label{ini}
\en
and require  $\chi(x,E)$  to be absolutely integrable on the full
real line; this restricts  the possible values taken by $E$ to a
discrete set. Fourier transforming (\ref{ini}) yields
\eq
\left( -{d^2 \over dk^2}-{ 1 \over k}{d \over dk} +( (-1)^n
k^{2n-2}-E) \right) \tilde{\chi}(k,E)=0~,
\en
and replacing
\eq
k \rightarrow i k~,~~E \rightarrow -E~,~~ \tilde{\chi}(k,E)
\rightarrow k^{-1/2} \tilde{\chi}(k,E)
\en
we finally find
\eq
\left( -{d^2 \over dk^2} + k^{2n-2}- { 1 \over 4 k^2} -E \right)
\tilde{\chi}(k,E)=0~.
\label{finalsh}
\en
Equation (\ref{finalsh})  exactly matches   the   ODE associated
in~\cite{Dorey:1998pt,Bazhanov:1998wj} to the reflectionless points
of the  untwisted sine-Gordon model at its $c{=}c_{\rm eff}{=}1$
conformal point.

This simple observation gives extra support to the correctness of
the  $D_n$ proposal (\ref{so2n2}), and it leads naturally to
the $B_n$ proposals discussed in the next section.

\resection{The $B_n$ models}
\label{secb}
The discussion in \S\ref{sgso2n} of the link  between the
sine-Gordon and $D_n$ scattering theories at  specific values  of the
parameters can be  extended to the  $B_n$
models~\cite{Tateo:1994pb,Dorey:1996ms}.
This and further considerations led us
to the ODE~(\ref{so2n10}), which we repeat here:
\eq
\left( D_{n}({\bf g^{\dagger}}) D_{n}({\bf g}) + \sqrt{P_K (x,E)}
\left(\frac{d}{dx} \right) \sqrt{P_K (x,E)} \right)\psi(x,E, {\bf
g})=0~.
\label{so2n101}
\en
The results of \S\ref{so2npsi} and~\S\ref{sd2}  suggest a link
between the presence of the integral operator $(d/dx)^{-1}$ and the
possibility of breaking  the symmetry between the  $n$ and $n{-}1$
nodes of the $D_n$ Dynkin diagram by choosing $\gamma_n \ne
\gamma_{n-1}$ in  the  BAE. Therefore, in writing (\ref{so2n101}) we
have omitted the integral operator $(d/dx)^{-1}$ of  (\ref{so2n2})
because, in contrast to the $D_n$ Dynkin diagrams, the $B_n$
diagrams  have no $\Z_2$ symmetry.

The relevant solution
$\psi(x,E,{\bf g})$ to (\ref{so2n101}) has the asymptotic $x \to\infty$
behaviour
\eq
\psi(x,E,{\bf g}) \sim   {\cal N} \; x^{-h^{\vee} M/2} \exp \left(-
{x^{M+1} \over M+1} \right)\quad,\quad (M >K/(h^{\vee}{-}K) )
\en
on the positive real axis.
The solutions~(\ref{chib})
\eq
\chi_i (x, E, {\bf g}) \quad ,\quad i = 0,1, \dots, 2n - 1
\en
are instead characterised by the $x \to 0$  behaviour
\eq
\chi_i(x, E, {\bf g}) \sim  x^{\lambda_i}+ O (x^{\lambda_i +
2n})~~~,~~~~ \left\{
\begin{array}{ll}
\lambda_i=g_i~, & {\rm for}~ i \le n-1~, \\
\lambda_i=h^{\vee}-g_{2n-1-i}~, & {\rm for}~ i>n-1~.
\end{array}
\right.
\label{lambdag1}
\en
In (\ref{lambdag1}) the  $\lambda$'s represent the $2n$ solutions of
the indicial equation in table~\ref{table2} with the ordering
\eq
~~~~g_i< g_j<h^{\vee}/2~,~~~ \lambda_i <
\lambda_j~,~~~~\forall  \ i<j~.
\en
\subsection{The $\psi$-system and the $B_n$ Bethe ansatz equations}
The $B_n$  $\psi$-system is
\eq
\psi^{(a-1)} \psi^{(a+1)}  = W[ \psi_{-\frac{1}{2}}^{(a)},
\psi_{\frac{1}{2}}^{(a)}]~~,~~a=1,\dots,n-1~,
\label{bnpsi1}
\en
\eq
\psi^{(n-1)}_{-\frac{1}{4}}  \psi^{(n-1)}_{\frac{1}{4}}  = W[
\psi_{-\frac{1}{4}}^{(n)}, \psi_{\frac{1}{4}}^{(n)}]~~.
\label{bnpsi2}
\en
Using  the  identity~(\ref{jacobi})   we can express $\psi^{(a)}$
with $a{>}1$ in terms of $\psi^{(1)}
\equiv \psi$. The result is
\eq
\psi^{(a)} =W[\psi_{\frac{1-a}{2}},\dots,
\psi_{\frac{a-1}{2}}]~,~~~a=1,2,\dots,n~.
\en

Following the derivation in  \S\ref{anBAe}, define
$\alpha_a = \sum_{i = 0}^{a - 1} \lambda_i$,
\eq
Q_k^{(a)} (E,{\bf g}) = Q^{(a)}_{[0, \dots, a - 1]} (\Omega^{k}
E,{\bf g})~,~~~ \bar{Q}_k^{(a)} (E,{\bf g}) = Q_{[0, \dots, a - 2,
a]}^{(a)} (\Omega^{k} E,{\bf g})
\en
where $Q^{(a)}_{[0, \dots, a - 1]}$  and $Q_{[0, \dots, a - 2,
a]}^{(a)}$ are defined as in (\ref{expanding}) and
\bea
\psi_k^{(a)} (x, E,{\bf g})& = &\omega^{k \alpha_a} Q_k^{(a)}(E,{\bf
g})
W^a [\chi_0, \dots, \chi_{a- 1}] + \nn \\
& &  \omega^{k (\alpha_{a + 1} -\lambda_{a - 1})} \bar{Q}_k^{(a)}
(E,{\bf g}) W^a [\chi_0, \dots, \chi_{a - 2}, \chi_a] +
\dots~. \nonumber\\
& & \label{psiexpansionbn}
\eea
Using (\ref{psiexpansionbn}) in (\ref{bnpsi1}) and  identifying the
leading contributions about $x{=}0$ gives
\eq
 Q^{(a + 1)} (E) Q^{(a - 1)} (E) =
   \omega^{\frac{1}{2} (\lambda_{a} - \lambda_{a-1})} Q_{- \frac{1}{2}}^{(a)}(E)
   \bar{Q}_{\frac{1}{2}}^{(a)}(E) - \omega^{\frac{1}{2} (\lambda_{a-1}-\lambda_{a} )}
   Q_{\frac{1}{2}}^{(a)} (E) \bar{Q}_{- \frac{1}{2}}^{(a)}(E)~
\en
with $Q^{(0)}(E)=1$ and
\eq
{Q^{(a- 1)}_{-\frac{1}{2}} (E^{(a)}_i) \over Q^{(a -
1)}_{\frac{1}{2}} (E^{(a)}_i)} {Q^{(a)}_{1}( E^{(a)}_i) \over
Q^{(a)}_{-1} (E^{(a)}_i)}{Q^{(a+ 1)}_{-\frac{1}{2}} ( E^{(a)}_i)
\over Q^{(a + 1)}_{\frac{1}{2}}(E^{(a)}_i)}
 =-
\Omega^{-2\gamma_a+\gamma_{a-1}+\gamma_{a+1}}~.
\label{BABn}
\en
In (\ref{BABn}),  $a=1,\dots,n-1$ and
\eq
\gamma_a= \alpha \left (\sum_{i=0}^{a-1} \lambda_i+ a v
\right)~,~~a=1,2,\dots,n~,
\en
where $\alpha={2K \over M h^{\vee}}$ and $v$ is still  to be fixed.
Plugging (\ref{psiexpansionbn}) into (\ref{bnpsi2})  leads to
\eq
 Q^{(n-1)}_{-\frac{1}{4}} (E) Q^{(n - 1)}_{\frac{1}{4}}  (E) =
   \omega^{\frac{1}{4} (\lambda_{n} -
\lambda_{n-1})} Q_{- \frac{1}{4}}^{(n)}(E)
   \bar{Q}_{\frac{1}{4}}^{(n)}(E) -
\omega^{\frac{1}{4} (\lambda_{n-1}-\lambda_{n} )}
   Q_{\frac{1}{4}}^{(n)} (E) \bar{Q}_{- \frac{1}{4}}^{(n)}(E)
\en
and
\eq
{Q^{(n- 1)}_{-\frac{1}{2}}(E^{(n)}_i) \over Q^{(n -
1)}_{\frac{1}{2}}( E^{(n)}_i)}
 {Q^{(n)}_{\frac{1}{2}}(E^{(n)}_i)
\over Q^{(n)}_{-\frac{1}{2}} ( E^{(n)}_i)} =- \Omega^{-\gamma_n+{1
\over 2}(\gamma_{n-1}+\gamma_{n+1})}
\label{BABn2}~.
\en
The boundary condition $\gamma_{n+1}{=}\gamma_{n-1}$ fixes
$v=-h^{\vee}/2$, the $\{ g_a \} \leftrightarrow \{ \gamma_a \}$
relation in table~\ref{table3} and allows (\ref{BABn}) and
(\ref{BABn2}) to be recast in the form~(\ref{dall0}).

We have  checked the consistency of the $n{=}2$ case both
numerically and analytically. The relation with the sine-Gordon
model briefly mentioned at the beginning of \S\ref{secb} and the
analysis  of \S\ref{b1} and \S\ref{b2=c2} lend extra analytic
support to the proposal.

\subsection{Example 3: $B_1$ }
\label{b1}
This is a singular limit for the analytic BAE study in
\cite{Kuniba:1994na}. It however suggests the equivalence of the
$K{=}1$ case of $B_1$ to the $K{=}2$ case of $A_1$ in the integrable
system. Indeed, the differential equation is second order in the
former case and it can be written in the more-standard form
\eq
\left(  {d^2 \over dx^2}
 + P_K(x,E) {d \over dx}+ {1 \over 2} \left({ d \over dx} P_K(x,E) \right) - {g_0(g_0-1) \over
x^2} \right) \psi(x,E,g_0)=0~.
\label{so3}
\en
Performing a Liouville transformation
\eq
\psi(x,E,g_0) \to  \psi(x,E,g_0) \exp \left(-{1 \over 2} \int^x
P_K(\xi,E) \,d\xi \right)
\en
we find
\eq
\left(  -{d^2 \over dx^2}
 + {1 \over 4} \left(x^{M/K}-E \right)^{2K} + {g_0(g_0-1) \over
x^2} \right) \psi(x,E,g_0)=0~.
\label{so31}
\en
Equation (\ref{so31}) coincides with the equations studied by Sergei
Lukyanov~\cite{Luk-private} which are related to the  $A_1$ lattice
models with integer spin
\eq
j=K~,~~~~K=1,2,3,\dots~.
\en

\begin{figure}[ht]
\begin{center}
\epsfxsize=.585\linewidth\epsfbox{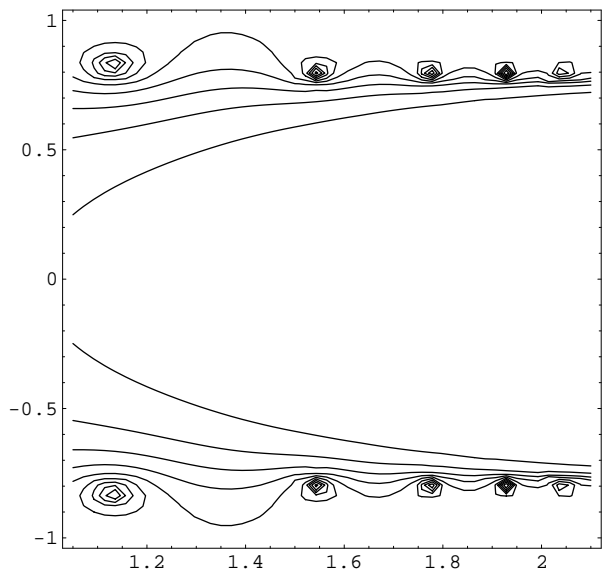}
\end{center}
\caption{  2-strings  in the $B_1$ model at $g_0{=}0, M{=}K{=}1$.
Contour plot of $1/(1+ |H_{-\frac{1}{2}} (-\fract{ 1}{\sqrt{2}}
e^{\te})|)$ in the complex $\te$-plane. } \label{B1fig}
\end{figure}

The cases $g_0{=}0$ with $M{=}K$ can be solved in closed form. After
a shift $x \rightarrow x+E$, the  simplest case $K{=}1$ becomes
\eq
\left( {d^2 \over dx^2}
 + x{d \over
dx}+\fract{1}{2} \right) \psi(x)=0
\label{lineraso3}
\en
which has  general solution
\eq
y(x)=c_1\;  e^{-{x^2 \over 2}} H_{-\frac{1}{2}}
\left(\fract{x}{\sqrt{2}}\right) +c_2 \;e^{-{x^2 \over 4}} \sqrt{x}
\; I_{-\frac{1}{4}}\left(\fract{x^2}{4} \right)
\en
where $H$ and $I$ are respectively the Hermite and the Bessel
functions. Since
\eq
H_{\nu}(x) \sim 2^{\nu} x^{\nu}(1+O(1/x))
\en
 as ${\rm Re}(x)>0$,
$|x| \to \infty$, while
\eq
I_{\nu}(x) \sim {e^{x} \over \sqrt{2 \pi x}} (1+ O(1/x))\,,
\en
 the most subdominant solution at large $x$ on the positive real
axis is
\eq
 \psi(x)= e^{-{x^2 \over 2}} H_{-\frac{1}{2}}
\left(\fract{x}{\sqrt{2}}\right)~.
\en
Figure~\ref{B1fig} shows the first five complex-conjugate pairs of
zeros of $\psi(-e^{\te})$. This  2-string pattern is typical of
$A_1$-related spin{-}1 integrable models. The exact eigenvalues are
reported in table~\ref{taba42} above.  There is also good agreement
between the position of the first pairs of  zeros shown in
figure~\ref{B1fig} and the WKB asymptotic prediction of
appendix~\ref{appa}.

\resection{The $C_n$ models}
\label{secc}
The analytic and numerical results of \S\ref{secd} support the
conjectured link between the $D_n$  BAE (\ref{dall0}) and
equation~(\ref{so2n2}). At  ${\bf g}{=}\{ 0,1,\dots,n-1 \}$, the
$D_n$ ODE is
\eq
\left( { d^{2n-1} \over dx^{2n-1}} - \sqrt{P_{K}(x,E)}
\left(\frac{d}{dx} \right) \sqrt{P_{K }(x,E)} \right)\psi(x,E)=0~.
\label{db00}
\en

In this section, we start from (\ref{db00}) and consider a second
 duality relation
\eq
{(\hat{D}_{-n})_K \times (\hat{D}_{-n})_L \over
 (\hat{D}_{-n})_{K+L}} \sim { (\hat{C}_{n})_{-K/2} \times
  (\hat{C}_{n})_{-L/2}
 \over  (\hat{C}_{n})_{-K/2 -L/2}}
\label{dualdc1}
\en
discussed by Hornfeck \cite{Hornfeck:1994is}. Motivated by the
results of \S\ref{negative} on  the analogy between the
$A_n \leftrightarrow A_{-n}$ spectral duality  and the  $A_n$-related
duality  in \cite{Hornfeck:1994is}, we change
$n \rightarrow -n$ and $K \rightarrow -2 K$ in (\ref{db00}):
\eq
\left((-1)^{-2n-1} { d^{-2n-1} \over dx^{-2n-1}} +\sqrt{P_{-2K
}(x,E)} \left(\frac{d}{dx} \right) \sqrt{P_{-2K} (x,E)}
\right)\psi(x,E)=0\,,
\en
where
\eq
P_{-2K}(x,E)= \left(x^{\frac{M(2n+2)}{2K}}-E \right)^{-2K}.
\en
Replacing
\eq
\psi(x,E) \rightarrow  \left
(\left(P_{-2K}(x,E)\right)^{-\fract{1}{2}} { \left(d \over dx
\right)^{-1}} \left(P_{-2K }(x,E)\right)^{-\fract{1}{2}}
\right)\psi(x,E)\,,
\en
multiplying by $d^{2n+1} \over dx^{2n+1}$ and noticing that
\eq
\left(P_{-2K}(x,E)\right)^{-\fract{1}{2}} \equiv P_K(x,E)=
(x^{h^{\vee} M/K} -E)^K\,,
\en
where $h^{\vee}{=}n+1$ is the dual Coxeter number of $C_n$, we can
write the resulting equation as
\eq
\left({d^{2n+1} \over dx^{2n+1}}  -P_{K}(x,E)  { \left(d \over dx
\right)^{-1}}P_{K} (x,E) \right)\psi(x,E)=0\,.
\label{finalcn}
\en
Equation (\ref{finalcn}) is our $C_n$ candidate at ${\bf g}{=}{\bf
g_0}{=}\{0,1,2,\dots \}$. Further, adapting the discussion of
\S\ref{secd} that led to the full $B_n$ equation, and
 noting  the similarity between
the pseudo-differential
operators (\ref{sun0}), (\ref{so2n0}) and (\ref{so2n10}),
we replace
\eq
{d^{2n+1} \over dx^{2n+1}} \equiv  D_{n}({\bf g^{\dagger}_0})
\left(\frac{d}{dx} \right) D_{n}({\bf g}_0) \Longrightarrow
D_{n}({\bf g^{\dagger}}) \left(\frac{d}{dx} \right)D_{n}({\bf g})~,
\en
and the  final $C_n$ proposal becomes (\ref{sp2n0}):
\eq
 \left( D_{n}({\bf g^{\dagger}}) \left(\frac{d}{dx} \right)D_{n}({\bf g})   -P_{K }(x,E)  { \left(d
\over dx \right)^{-1}}P_{K} (x,E) \right)\psi(x,E, {\bf g})=0~.
\label{sp2n1}
\en\\
The relevant solution of (\ref{sp2n1}) has the asymptotic $x
\to\infty$ behaviour
\eq
\psi(x,E,{\bf g}) \sim   {\cal N} \; x^{-n M} \exp \left(- {x^{M+1}
\over M+1} \right)\quad , \quad (M >K/(h^{\vee}{-}K) )
\en
on the positive real axis.
The solutions~(\ref{chib})
\eq
\chi_i (x, E, {\bf g})~,~~~ i = 0,1, \dots, 2n + 1
\en
are characterised by the $x \to 0$  behaviour
\eq
\chi_i(x, E, {\bf g}) \sim  x^{\lambda_i}+ O (x^{\lambda_i +
2n+1})~~~,~~~~ \left\{
\begin{array}{ll}
\lambda_i=g_i~, & {\rm for}~ i \le n-1, \\
\lambda_n=n~,   &~~~~~~~~~~~~ \\
\lambda_i=2n-g_{2n-i}~, & {\rm for}~ i>n~.
\end{array}
\right.
\label{lambdag3}
\en
In (\ref{lambdag3}) the  $\lambda$'s represent the $2n+1$ roots of
the indicial equation in table~\ref{table2} with the ordering
\eq
~~~~g_i< g_j<n~,~~~ \lambda_i < \lambda_j~,~~~~\forall \ i<j~.
\en
\subsection{The $\psi$-system and the $C_n$ Bethe ansatz equations}
We now deduce the $C_n$ Bethe ansatz equations from the proposed
 $C_n$ $\psi$-system:
\bea
W[\psi^{(a)}_{-\frac{1}{4}}, \psi^{(a)}_{\frac{1}{4}}] &=&
\psi^{(a-1)} \psi^{(a+1)}~,~~~~ a=1,2,\dots,n-2~, \nn\\
W[\psi^{(n-1)}_{-\frac{1}{4}}, \psi^{(n-1)}_{\frac{1}{4}}] &=&
\psi^{(n-2)} \psi^{(n)}_{-\frac{1}{4}}  \psi^{(n)}_{\frac{1}{4}}~,
\label{wronskianCnlastm1} \\
W[\psi^{(n)}_{-\frac{1}{2}}, \psi^{(n)}_{\frac{1}{2}}] &=&
\psi^{(n-1)}~.
\nn
\eea
Using the Jacobi identity we find
\eq
\phi^{(a)} \equiv \psi^{(a)}
=W^{(a)}[\psi_{\frac{1-a}{4}},\psi_{\frac{3-a}{4}}\dots,
\psi_{\frac{a-1}{4}}]~,~~~a=1,2,\dots,n-1~,
\en
\eq
\phi^{(n)} \equiv \psi^{(n)}_{-\frac{1}{4}} \psi^{(n)}_{\frac{1}{4}}
=W^{(n)}[\psi_{\frac{1-n}{4}},\psi_{\frac{3-n}{4}},\dots,
\psi_{\frac{n-1}{4}}]~,
\en
where the  functions $\phi^{(a)}(x,E,{\bf g})$ satisfy the following
system of  functional relations
\eq
W[ \phi^{(a)}_{-\frac{1}{4}} ,\phi^{(a)}_{\frac{1}{4}} ]
=\phi^{(a-1)} \phi^{(a+1)}~,~~~a=1,\dots, n-1~,
\label{antype1}
\en
and
\eq
W[ \phi^{(n)}_{-\frac{1}{4}} ,\phi^{(n)}_{\frac{1}{4}} ]
=\phi^{(n-1)} (\psi^{(n)})^2~.
\label{expsi}
\en
{}From (\ref{expsi}) we see that
\eq
\psi^{(n)}(x,E,{\bf g})= \sqrt{ W^{(n+1)}[\psi_{-\frac{n}{4}},\dots,
\psi_{\frac{n}{4}}]}~.
\en
Now set
\bea
\phi_k^{(a)} (x, E,{\bf g})& =
 &\omega^{k \alpha_a} \hat{Q}_k^{(a)}(E,{\bf g}) W^{(a)} [\chi_0, \dots, \chi_{a- 1}] + \nn \\
& &  \!\! \!\! \omega^{k (\alpha_{a + 1} -\lambda_{a - 1})} \bar{Q}_k^{(a)}
(E,{\bf g}) W^{(a)} [\chi_0, \dots, \chi_{a - 2}, \chi_a] + \dots
\label{above3}
\eea
with $a=1,2,\dots,n$. The  orders of the first and the second term
in (\ref{above3}) are given respectively by
$\alpha_a - a (a - 1) / 2$ and
$\alpha_{a + 1} - \lambda_{a - 1} - a (a - 1) / 2$, where $\alpha_a =
\sum_{i = 0}^{a - 1} \lambda_i$.
Assuming
\eq
\psi_k^{(n)} (x, E,{\bf g})=\omega^{k \beta} x^\beta
Q_k^{(n)}(E,{\bf g})+\omega^{k \sigma} x^\sigma
\Tilde{Q}_k^{(n)}(E,{\bf g})+\dots ~
\en
where $\beta<\sigma$, we can make the following identifications
\eq
\beta= \alpha_n/2
-n(n-1)/4~~,~~\sigma=\beta+(\lambda_n-\lambda_{n-1})
\en
and
\bea
\hat{Q}^{(0)}(E) &=& Q^{(0)}(E)=1~,\\
 \hat{Q}^{(a)}(E) &=&
Q^{(a)}(E)~,~~~a=1,2,\dots,n-1 \label{DDQ1}~, \\
\hat{Q}^{(n)}(E) &\propto& Q^{(n)}_{-\frac{1}{4}}(E)\;
Q^{(n)}_{\frac{1}{4}}(E)~~ \label{DDQ2}~, \\
\bar{Q}^{(n)}(E) &\propto&
\omega^{\frac{1}{4}(\lambda_n-\lambda_{n-1})}Q^{(n)}_{-\frac{1}{4}}
(E)\; \tilde{Q}^{(n)}_{\frac{1}{4}} (E) +
\omega^{\frac{1}{4}(\lambda_{n-1}-\lambda_{n})}
Q^{(n)}_{\frac{1}{4}}
(E)\; \tilde{Q}^{(n)}_{-\frac{1}{4}}(E).~ \nn\\
 ~~\label{DDQ3}
\eea
Using (\ref{above3}) in (\ref{antype1}) and selecting the leading
terms we find
\eq
 \hat{Q}^{(a + 1)} (E) \hat{Q}^{(a - 1)} (E) =
   \omega^{\frac{1}{4}
(\lambda_{a} - \lambda_{a-1})} \hat{Q}_{- \frac{1}{4}}^{(a)}(E)
   \bar{Q}_{\frac{1}{4}}^{(a)}(E) -
 \omega^{\frac{1}{4} (\lambda_{a-1}-\lambda_{a} )}
   \hat{Q}_{\frac{1}{4}}^{(a)} (E) \bar{Q}_{- \frac{1}{4}}^{(a)}(E)\,.
\en
With the identifications (\ref{DDQ1}) and (\ref{DDQ2}) this leads to
\eq
\prod_{ b=1}^{n} \frac {Q^{(b)}_{B_{ab}}(E^{(a)}_{i})}
{Q^{(b)}_{-B_{ab}}(E^{(a)}_{i})}= - \Omega^{ {\alpha \over 4}
(\lambda_a-\lambda_{a-1})}\,,~
\label{dall000}
\en
where $a=1,2,\dots, n-1$ and  $\alpha={2 K \over M h^{\vee}}$\,.
Using (\ref{above3}) with $a{=}n$ in (\ref{expsi}), (\ref{DDQ3}) and
(\ref{DDQ2}) gives instead
\eq
Q^{(n -1)} (E)= \omega^{\frac{1}{2}
(\lambda_{n } - \lambda_{n-1})} Q_{- \frac{1}{2}}^{(n)}(E)
   \tilde{Q}_{\frac{1}{2}}^{(n)}(E) -
\omega^{\frac{1}{2} (\lambda_{n-1}-\lambda_{n} )}
   Q_{\frac{1}{2}}^{(n)} (E) \tilde{Q}_{- \frac{1}{2}}^{(n)}(E);
\en
which leads to
\eq
{Q^{(n- 1)}_{-\frac{1}{2}}  ( E^{(n)}_i) \over Q^{(n -
1)}_{\frac{1}{2}} (E^{(n)}_i)}
 {Q^{(n)}_1 ( E^{(n)}_i)
    \over
 Q^{(n)}_{-1} (E^{(n)}_i)}
=- \Omega^{ {\alpha \over 2} (\lambda_n-\lambda_{n-1} )}~.
\label{fffBA1}
\en
Finally, with the identification (\ref{lambdag3}) and the choice in
table \ref{table3} for the $\{   g_a \} \leftrightarrow  \{
{\gamma_a} \}$ relation, equations (\ref{dall000}) and
(\ref{fffBA1}) can be assembled into the universal form (\ref{dall0}).
\subsubsection{Example 4: $C_1$}
The $n{=}1$ case is
 again a singular limit of the analytic BAE,
but it also suggests the similarity of this case to the $A_1$ models
\cite{Kuniba:1994na}. The pseudo-differential equation is, however,
not second order but instead third order
\eq
\left( {d^3 \over dx^3}-{{\cal L} \over x^2} {d \over dx}   + {{\cal
L} \over x^3} - P_K(x,E) \left(d \over dx \right)^{-1} P_K(x,E)
\right) \psi(x,E,g_0)=0
\label{ps1}
\en
where ${\cal L}=g_0(g_0-2)$. It is nevertheless  easy to check that
(\ref{ps1}) is solved by the product of two functions satisfying
second order ODEs:
\eq
\psi(x,E,g_0)=\chi_-(x,E,g_0) \chi_+(x,E,g_0)
\en
where $\chi_{\pm}$ originates from a single function $\chi$ as
follows
\eq
\chi_{\pm}(x,E,g_0)= \chi( \omega^{\pm 1/4}x, \Omega^{\pm 1/4} E,
g_0).
\label{def4}
\en
Since we assume that
$\chi$ satisfies the standard ODE associated with $A_1$\,,
\eq
\left ({d^2 \over dx^2} -  {1 \over 2} P_K(x,E) - {{\cal L} \over 4
x^2} \right) \chi(x,E,g_0)=0\,,
\en
the functions $\chi_{\pm}$ satisfy the following:
\eq
{ d^2 \over dx^2} \chi_{\pm}(x,E,g_0)= \left(\pm { i \over 2}
P_K(x,E)+ { {\cal L}\over 4x^2} \right) \chi_{\pm}(x,E,g_0)\,.
\en
Starting from (\ref{def4}) and differentiating
$\psi=\psi(x,E,g_0)$ three times we find
\bea
{d\psi \over dx}&=&{d \chi_+\over dx} \; \chi_-+\chi_+{d \chi_-\over
dx}\,;
{}~~~ {d^2 \psi\over dx^2} ~=~2 {d \chi_-\over dx}  {d \chi_+\over dx} +
{{\cal L} \over 2 x^2} \psi~; \nn \\
{d^3 \psi\over dx^3}&=&i P_K \; (\chi_+ {d \chi_-\over dx}  - {d
\chi_+\over dx} \chi_-) + {{\cal L} \over x^2} {d\psi \over dx} -
{{\cal L} \over x^3} \psi\,.
\label{fn1}
\eea
We notice that
\eq
{d \over dx}\lf (\chi_+ {d \chi_- \over dx}- {d\chi_+ \over dx} \chi_-\ri)
=  -i P_K \; \psi
\en
and therefore
\eq
\chi_+ {d \chi_-\over dx} - {d \chi_+\over dx} \chi_-= -i \left(d
\over dx \right)^{-1} P_K  \psi.
\label{lat1}
\en
Inserting  (\ref{lat1}) into  (\ref{fn1})  we  finally arrive at
equation (\ref{ps1}).
\subsubsection{Example 5: $C_2 \sim B_2$}
\label{b2=c2}
The ODEs for $B_2$ and for $C_2$ are both deduced from the
$D_n$-type ODE, but through completely different routes. It is thus
a good test to check the equivalence of these two cases. We start
with the $B_2$-related ODE at ${\bf g}{=} \{0,1 \}$
\eq
{d^{4}\psi \over dx^4} +P_K {d\psi \over dx} +{1 \over 2} {dP_K
\over dx}\psi=0.
\en
The ODE associated with the second node of $B_2$, which is nothing
but the first node of $C_2$, would be satisfied by the following
function
\eq
\psi^{(2)}=W[ \psi_{-\frac{1}{2}}, \psi_{\frac{1}{2}}]=[0,1]~,
\en
where
\eq
{d^{4}\psi_{\pm \frac{1}{2}} \over dx^4}=  P_K {d\psi_{\pm
\frac{1}{2}} \over dx} +{1 \over 2} {dP_K \over dx}\psi_{\pm
\frac{1}{2}}~.
\en
We then easily evaluate the derivatives of $\psi^{(2)}$:
\bea
{d\psi^{(2)} \over dx} &=&[0,2]~,~~{d^2\psi^{(2)} \over dx^2}=[1,2]+[0,3]~,~~ {d^3\psi^{(2)} \over dx^3}=2[1,3] + P_K \psi^{(2)} \nn\\
 \nn \\
{d^4\psi^{(2)} \over dx^4}&=&2\; [2,3]+P_K {d\psi^{(2)} \over
dx}~,~~{d^5\psi^{(2)} \over dx^5} =P_K{d^2\psi^{(2)} \over dx^2}  -
2P_K \;[1,2]~.
\eea
Finally, noticing that
\eq
2[1,2]= 2  \left({d \over dx}\right)^{-1}[1,3]={d^2\psi^{(2)} \over
dx^2}- \left({d \over dx}\right)^{-1} P_K \psi^{(2)}
\en
we have obtained a closed form ODE for $\psi^{(2)}$:
\eq
\left({d^5\psi^{(2)} \over dx^5} - P_K(x,E) \left({d \over
dx}\right)^{-1} P_K(x,E) \right)\psi^{(2)}(x,E)=0~.
\en
This equation  is exactly  the ${\bf g}{=}\{0,1 \}$ equation
associated to the first node of $C_2$. This verifies the consistency
of our proposal.

More generally, the correspondence between the $B_2$ parameters
${\bf g}{=}\{g_0, g_1 \}$ and the $C_2$ parameters ${\bf
\bar{g}}{=}\{\bar{g}_0, \bar{g}_1 \}$ is established by
\bea
2g_0 &=&\bar{g}_0+\bar{g}_1-1~, \nn \\
2g_1  &=&\bar{g}_0-\bar{g}_1+3~.
\eea

Finally we have directly checked the $C_2{=}B_2$  $\psi$-system at
${\bf g}{=}\{0,1 \}$ by verifying that both sides of

\eq
W[ \psi^{(1)}_{-\frac{1}{4}} ,\psi^{(1)}_{\frac{1}{4}} ]
=\psi^{(2)}_{-\frac{1}{4}} \psi^{(2)}_{\frac{1}{4}}~,
\en
 satisfy the same 15th order ODE.

Unfortunately we do not currently have NLIEs for the $B_n/C_n$ models and the
numerical checks presented below are instead based  on an
approximate  solution similar to that used by Voros in \cite{Voros}.
For the  $B_2$
BAEs for the case $ {\bf g}{=}\{0,1\}$, $K{=}1$ and
$M{=}2/3$, we started from a perfect-string estimate for the first 1000
roots as follows:
\begin{align}
E_j &= \bigl (  4\sqrt{\frac{\pi}{3}}
 \frac{\Gamma(\frac{11}{6})}{\Gamma(\frac{4}{3})} j \bigr)^{\frac{6}{5}}
 &\text{1st node} \\
E_j &= {\rm e}^{\pm i \frac{\pi}{5}}
\bigl ( 4 \sqrt{\pi} \frac{\Gamma(\frac{11}{6})}{\Gamma(\frac{4}{3})}
 (j-\frac{1}{6}) \bigr)^{\frac{6}{5}}
 &\text{2nd node.}
\label{twostr}
\end{align}

\begin{table}[ht]
\begin{center}
\begin{tabular}{|l  l | l   l|}
\hline ~BA~numerics& &~ODE numerics & {}   \\
1st~node~~~~&2nd~node~~~~&1st~node~~~~&2nd~node~~~~   \\
\hline   6.28405 & $6.8368 \pm  5.8640i$ & 6.28390 &$6.8365 \pm
5.8637i$~  \\
        13.2379  & $18.216 \pm 14.265i$ & 13.2376 &$18.214\pm
14.264i$~    \\
        21.6307  & $30.996 \pm 23.645i$  & 21.6303 &$30.992 \pm
23.642i$~  \\
        30.5039 & $44.747 \pm 33.707i$  & 30.5034&$44.739\pm 33.700i$~
\\
        39.8617 & $59.254 \pm 44.304i$  & 39.8613&$59.240 \pm
44.292i$~   \\
\hline \end{tabular} \caption{\footnotesize $C_2{=}B_2$: comparison of
Bethe ansatz results
with the numerical solution of the $B_2$ and $C_2$ equations using
the algorithm described in appendix~\ref{appb}. ($B_2$ node
convention with ${\bf g}{=}\{0,1\}$, $K{=}1$ and
$M{=}2/3$.)\label{tabc2b2}}
\end{center}
\end{table}

We then solved the BAEs recursively using the  Newton-Raphson method
on the first 20 roots, keeping the remaining roots fixed.
Table~\ref{tabc2b2} compares the lowest roots thus obtained with the
results from the solution of the (pseudo-)differential equations.
The relatively low accuracy in comparison with tables~\ref{taba4},
\ref{taba42} and \ref{tabd4} is most likely to be a consequence of
the slow convergence rate of the algorithm used to solve the Bethe
ansatz equations, and in particular the systematic errors introduced
by fixing the higher levels ($E_j$, $j > 20$) to their
perfect-string values.
\section{Conclusions}
\label{conclusions}
There are many aspects about the correspondence between  integrable
models and the spectral theory of ordinary differential equations
that we would like to explore and understand at a deeper level.
Given the  applications of the Bethe ansatz to the study of
QCD in its leading-logarithm approximation~\cite{Lipatov:1994xy} and
to the study of anomalous dimensions of composite operators in
Yang-Mills theories~\cite{Minahan:2002ve,Ferretti:2004ba}, the
extension of the correspondence to lattice models is certainly
desirable from a physical point of view. On the other hand the
mathematical structures arising from the generalisation of the
correspondence to other conformal field theories with extended
symmetries has the potential to link areas of modern and classical
mathematics in an elegant way. In this paper our results were
obtained very much on a case-by-case basis, but we already saw the
emergence of interesting mathematical objects: the $\psi$-systems,
negative-dimension dualities, and the formal similarity with the
Miura-opers studied both in the classical
work~\cite{Drinfeld:1984qv} (see also \cite{Gelfand:1975rn,
Balog:1990mu, DiFrancesco:1990qr})  and more recently
in~\cite{Mukhin:2002fp, Mukhin:2002fp1, Mukhin:2002fp2,
Frenkel:2005fr, Chervov:2006xk}.

It would be very interesting to generalise equations
(\ref{sun0}--\ref{sp2n0}) to encompass the excited states of the
integrable models; to date this has been completed  for the
$K{=}1$ case of $A_1$~\cite{Bazhanov:2003ni, Fioravanti:2004cz}.
More challenging, but also extremely interesting, would be  to
extend the correspondence to perturbed conformal field theories
defined on a cylinder, both for the ground
state~\cite{Zamolodchikov:1989cf} and for excited
states~\cite{Bazhanov:1996aq, Dorey:1996re}.

Finally, even remaining inside the current setup, the ODE/IM
correspondence has already had an impact
on condensed matter physics:  it has been applied to interacting Bose
liquids~\cite{Gritsev:2006}, the single electron
box~\cite{Lukyanov:2006cu} and  quantum dots~\cite{Bazhanov:2003ua}.

\medskip
\noindent{\bf Acknowledgements --}
We are very grateful  to  Herman Boos, Boris Dubrovin, Ludwig
Faddeev, Frank Goehmann, Andreas Kl{\"u}mper, Sergei Lukyanov and
Fedor Smirnov for useful conversations and kind encouragement. PED,
TCD and JS thank  Torino University for hospitality at beginning of
this project. JS also thanks the members of the Universities of
Wuppertal and Bologna for hospitality and the Ministry of Education
of Japan for a `Grant-in-aid for Scientific Research', grant number
17540354. This project was also partially supported by the European
network EUCLID (HPRN-CT-2002-00325), INFN grant TO12, NATO grant
number PST.CLG.980424 and  The Nuffield Foundation grant number
NAL/32601, and a grant from the Leverhulme Trust.
%

%
%
\appendix
\resection{Root strings from the complex WKB method}
\label{appa}
As mentioned in \S\ref{BAe}, a characteristic feature of the `fused'
models with $K{>}1$ is that the asymptotic roots are not necessarily
real, but rather are grouped into so-called `strings' of complex
roots. Here we show how this behaviour can be recovered from a
treatment of the ODE using the complex WKB method. We shall restrict
the analysis to the $A_1$ case and for simplicity  set ${\bf
g}{=}\{0,1 \}$. The ODE is the $n{=}2$ case of (\ref{gnde}):
\eq
\left(-\frac{d^2}{dz^2}+P_K(z)\right)\psi(z,E)=0\,,\quad
P_K(z)=(z^{2M/K}{-}E)^K
\label{kode}
\en
and the boundary condition determining the eigenvalues is that there
should be a solution which decays as $z\to\infty$ on the positive
real axis, and is simultaneously zero at $z{=}0$. (In this section
we use $z$ instead of $x$ to emphasise that it must be considered in
the complex plane.)

\begin{figure}[ht]
\begin{center}
\epsfxsize=.575\linewidth\epsfbox{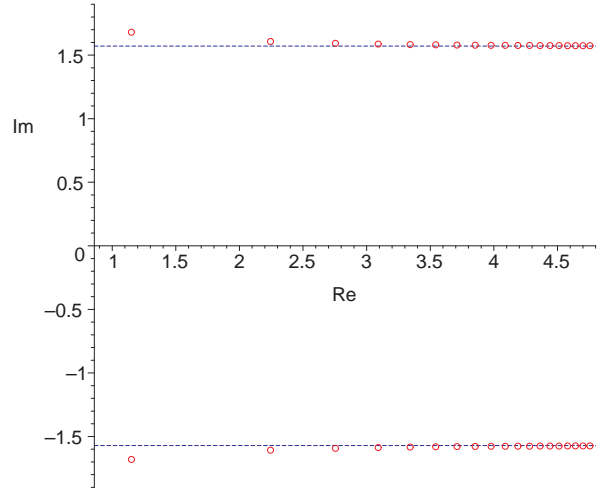}
\end{center}
\caption{Two-strings from the asymptotics of the NLIE~\cite{KBP}.}
\label{twostrings}
\end{figure}

Before giving the WKB treatment, we summarise the expectations from
integrable models. A useful discussion of the $K{=}2$ case, for
generic twist parameter,  is in appendix 2 of \cite{KBP}. For a
finite lattice of size $N$, the integral equation derived in that
paper leads to the following asymptotic condition on the Bethe
ansatz roots $\theta^{\pm}_j$
\eq
N\ln\left(\tanh\left(\frac{\pi\theta_j^{\pm}}{2\gamma}\right)\right)
=\pm N \pi i\mp (2j{-}1 + \frac{\phi}{\pi{-}2\gamma})\pi i +
\frac{1}{2}\ln 2~, \label{lattice}
\en
where $1\leq j<N/2+1$. This equation describes the approximate
position of complex-conjugate pairs of roots $\{ \te^-_j, \te^+_j
\}$, with $\te^{-}_j$  and  $\te^{+}_j$ lying in the upper and lower
halves of the complex plane.
  The relationship between
$\gamma$, $\phi$ in  (\ref{lattice}) and $M$ and ${\bf g}{=}\{g_0,
1-g_0 \}$ is
\eq
\mu={M+1 \over M}={\pi \over 2 \gamma}~,~~~ {\phi \over \pi -2
\gamma}= {1 \over 2}-g_0~.
\en
In the continuum limit, $N \to \infty$ and the numbers
$e^{\theta_j^{\pm}}$ with
$1 \leq j\ll N/2$ tend to zero. Shifting
$\theta_j^{\pm}$ appropriately, equation (\ref{lattice}) becomes
\eq
e^{2 \mu \theta_j^{\pm}}=\pm (2j - \frac{1}{2} - g_0)\pi i
-\frac{1}{2}\ln 2~,~~j=1,2,\dots~. \label{WKBs2}
\en
The second term causes deviations from the string pattern, even
within this asymptotic approximation.  Set
$ \theta_j^{\pm}=\pm i\pi/(4 \mu)+ \beta_j^{\pm}$\,; then
\eq
e^{2 \mu \beta_j^{\pm}}=(2j - {1 \over 2} - g_0)\pi \pm
\frac{i}{2}\ln 2\,
\en
or, taking logs and expanding for large $j$,
\eq
2 \mu \beta_j^{\pm}= \ln(2j - \frac{1}{2}-g_0)\pi \pm i\frac{\ln
2}{2(2j - \frac{1}{2}-g_0)\pi}+\dots~. \label{shift}
\en
Equation (\ref{shift}) exhibits the asymptotic deviations from the
perfect string configurations, which are only recovered in the limit
$j\to\infty$. This qualitative pattern matches the results of
table~\ref{taba42} and figure~\ref{B1fig}, and is illustrated in
figure~\ref{twostrings}. To compare with the  $A_1$, $K{=}2$ results
in \S\ref{anBAe} one has to identify
\eq
\{e^{2 \mu \te_j^{+}}, e^{2 \mu \te_j^{-}}  \} = \{ \mu^{-1}
(E_{2j-1})^{\mu},  \mu^{-1} (E_{2j})^{\mu} \}~
\en
in (\ref{WKBs2}).

These features (and their generalisations to higher $K$) can all be
recovered from a complex WKB treatment of (\ref{kode}). The
practicalities of this method have been very clearly explained by
Heading \cite{headingbook,headingglobal}, and we shall,
more-or-less, follow his notations, which we now summarise. The
behaviour of (\ref{kode}) is controlled by
$P_K(z)\equiv (z^{2M/K}-E)^K$, and the possibly-complex zeros of this
function are called turning (or transition) points. For each turning
point $z_0$, a branch cut is inserted joining
$z_0$ to infinity; and for $2M/K\notin\Z$, another is added starting from the origin along
the negative-imaginary axis. After this, taking $\arg\left( P_K(z)
\right)$ to tend to zero along the positive real $z$ axis renders
$(P_K(z))^{1/2}$ and
$( P_K(z))^{1/4}$ uniquely defined.
{}From each turning point $z_0$ there is a set of lines
\eq
\Im m\, \int_{z_0}^{z} dt \sqrt{P_K(t)} =0
\en
called Stokes lines, and another set
\eq
\Re e\, \int_{z_0}^{z} dt \sqrt{P_K(t)} =0
\en
called anti-Stokes lines\footnote{Here we follow the conventions of
Heading and, for example, Berry, but beware that other's conventions
are exactly the reverse. Note also that our $P_K(z)$ is {\em
minus}\/ the function $q(z)$ used by Heading.}. A $K^{\rm th}$ order
turning point has $K{+}2$ Stokes lines, and $K{+}2$ anti-Stokes
lines, emanating from it. Leading-order WKB solutions are written as
\eq
(z_0,z)\equiv (P_K(z))^{-1/4}\exp\left(\int_{z_0}^z dt \sqrt{P_K(t)}
\right)~.
\en
If $\Re e \int_{z_0}^z dt \sqrt{P_K(t)} >0$, a subscript $d$ is
added, signifying that $(z_0,z)_d$ is a dominant solution;
$(z,z_0)$ is then subdominant, and is written $(z,z_0)_s$.
On anti-Stokes lines, dominant and subdominant solutions swap
roles\footnote{Note, in contrast to elsewhere in this paper, the
terms dominant and subdominant are used here with respect to the WKB
expansion parameter, which for brevity we have set equal to $1$. The
parameter, a factor of $\varepsilon^{2}$ in front of the derivative
term in (\ref{kode}), can easily be restored if desired. In
Heading's notations, $\varepsilon$ corresponds to $k^{-1}$.}.

Formal WKB solutions $\alpha\,(z,z_0)+\beta\,(z_0,z)$ are only valid
in restricted domains: the Stokes phenomenon means that the
coefficient of the subdominant term changes by an amount
proportional to the dominant term when a Stokes line is crossed, so
that $\alpha\,(z,z_0)_d+\beta\,(z_0,z)_s$ is replaced by
$\alpha\,(z,z_0)_d+(\beta+T\alpha)\,(z_0,z)_s$\,, where the Stokes
multiplier $T$ is a constant characteristic of the given Stokes line
and crossing direction. (Note, since the discontinuity always occurs
in the coefficient of a subdominant term, there is no contradiction
with the fact that the WKB solution provides an asymptotic
approximation to the exact -- continuous -- solutions of the
original equation.) For a Stokes line emanating from a $K^{\rm th}$
order turning point, traversed in a positive (anticlockwise) sense,
the Stokes multiplier is~\cite{headingstokes}
\eq
T=2i\cos(\pi/(K{+}2))\,.
\en
This can alternatively be recovered via the $E{=}0$, $l{=}0$,
$M{=}K/2$ value of the function $C(E,l)$ discussed in
\cite{Dorey:1999uk}, equation (3.10). For a clockwise traverse, $T$
is replaced by $-T$.

The key feature of (\ref{kode}), leading to the formation of
strings, is that the turning points which control the WKB
eigenvalues all have order $K$. If a solution which decays along the
positive real axis is traced in to a turning point, there are $K$
different directions away from that turning point along which the
two WKB solutions $(z_0,z)$ and $(z,z_0)$ might come into balance
and have a chance to cancel; if the continuation of any of these
directions can be arranged to hit the origin, then a (WKB)
eigenvalue is possible. When the Stokes multipliers have unit
modulus, these directions are precisely the anti-Stokes lines;
departures from unit modulus cause small shifts, and lead to
deviations from the string pattern.

\begin{figure}[ht]
\begin{center}
\epsfxsize=.585\linewidth\epsfbox{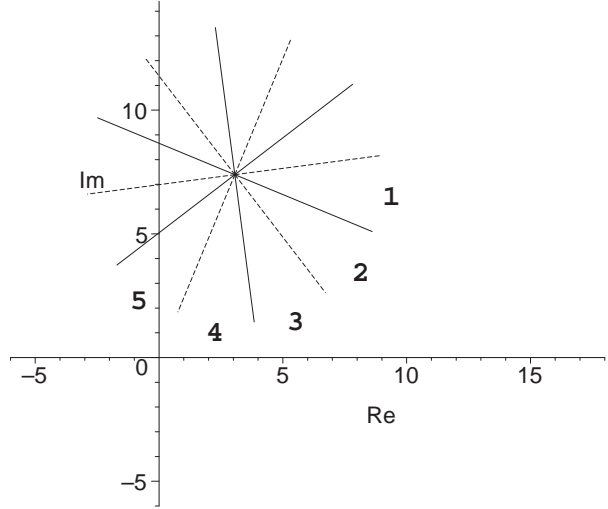}
\end{center}
\caption{Initial directions of Stokes lines (continuous) and anti-Stokes
lines (dotted) from a $4^{\rm th}$ order turning point.}
\label{stokesdirections}
\end{figure}

\begin{figure}[ht]
\begin{center}
\epsfxsize=.585\linewidth\epsfbox{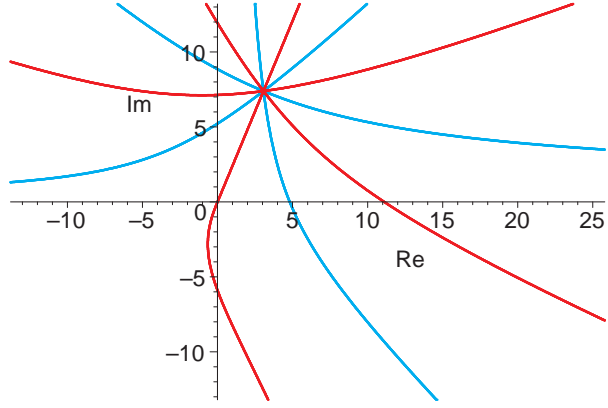}
\end{center}
\caption{The full Stokes and anti-Stokes lines for the situation
shown in figure~\protect\ref{stokesdirections}.}
\label{stokeslines}
\end{figure}

An example should make this clear. Figure \ref{stokesdirections}
shows the set of Stokes and anti-Stokes directions away from a
turning point of (\ref{kode}) with $M{=}3$, $K{=}4$. Between each
pair of anti-Stokes lines there is a Stokes sector, which is further
divided into two segments by the Stokes line which lies exactly in
the middle of the Stokes sector. The first few of these segments are
labelled {\bf 1} $\dots$ {\bf 5} in the figure; note that segments
{\bf 1} and {\bf 2} make up one Stokes sector, and segments {\bf 3}
and {\bf 4} another. For the figure, $\arg(E)$ was chosen so that
the origin lies in the union of segments  {\bf 4} and {\bf 5}, and
the positive real axis in the union of the continuations of segments
{\bf 1} and {\bf 2}. Across the (dotted) anti-Stokes lines, dominant
and subdominant solutions swap over, while across the (continuous)
Stokes lines, the coefficients of subdominant terms may change by
the Stokes phenomenon. Figure \ref{stokeslines} shows the
continuation of the Stokes directions shown in figure
\ref{stokesdirections} into the full complex plane, justifying the
claim that the real axis lies in the continuation of segments {\bf
1} and {\bf 2}.

Our aim is to find an approximate
solution which decays as $z\to\infty$ on the
positive real axis, and is also zero at the origin.
Since the positive real axis lies in the continuations of segments
{\bf 1} and {\bf 2}, we start with a solution subdominant
in segments {\bf 1} and {\bf 2}, and continue
it clockwise around the turning point to find the behaviour in the
region of the origin. There is no Stokes phenomenon passing between
{\bf 1} to {\bf 2} since the dominant term is absent in both of these
segments, but from then on the phenomenon must be taken into account
each time a Stokes line is crossed. With $z_0$ the location of the
turning point, and
$(z,z_0)$ the approximate subdominant solution in
segments {\bf 1} and {\bf 2}, we have:

\begin{tabbing}
\quad\= {\bf 1}, {\bf 2} \=:\quad $(z,z_0)_s$\\
\> ~~{\bf 3} \>:\quad $(z,z_0)_d$\\
\> ~~{\bf 4} \>:\quad $(z,z_0)_d-2i\cos(\pi/6)(z_0,z)_s$\\
\> ~~{\bf 5} \>:\quad $(z,z_0)_s-2i\cos(\pi/6)(z_0,z)_d$\\
\end{tabbing}

Near to the anti-Stokes line between segments {\bf 4} and {\bf 5},
the WKB solutions $(z,z_0)$ and $2i\cos(\pi/6)\,(z_0,z)$ are of
similar magnitude, and the (WKB) condition yielding the eigenvalues
is found by demanding that they cancel exactly at $z{=}0$:
\eq
(z_0,0)-2i\cos(\pi/6)(0,z_0)=0\,.
\en
That is
\eq
\exp\left(2\int_0^{z_0} dt
\sqrt{P_K(t)}+i\pi/2\right)=\frac{1}{2\cos(\pi/6)}
\label{dev}
\en
or
\eq
2\int_0^{z_0} dt \sqrt{P_K(t)}= (2j{-}\frac{1}{2})\pi i
-\ln(2\cos(\pi/6))~. \label{logdev}
\en
Remembering that $z_0{=}E^{K/(2M)}$, the integral can be evaluated
to
\bea
\int_0^{E^{K/(2M)}}\!\! (t^{2M/K}{-}E)^{K/2}\,dt&=&
(-1)^{K/2}E^{\mu}\int_0^1(1-u^{2M/K})^{K/2}\,du\nn\\
&=&(-1)^{K/2}\,\tilde\kappa(2M/K,2/K)\,E^{\mu}
\eea
where $\mu=K(M{+}1)/(2M)$ and
\eq
\tilde\kappa(a,b)=
\frac{\Gamma(1+1/a)\Gamma(1+1/b)}{\Gamma(1+1/a+1/b)}\,. \label{tk}
\en
Equations (\ref{logdev}---\ref{tk}) give the  WKB prediction at
$K{=}4$. Had we instead performed the calculation for $K{=}2$, the
logarithm on the RHS of (\ref{logdev}) would have been replaced by
$-\ln(2\cos(\pi/4))=-\frac{1}{2}\ln 2$, and (\ref{WKBs2}) would have
been reproduced. More generally, the Stokes multipliers solve a
stationary T-system, and the constant determining the deviation of
the $j^{\rm th}$ root in a $K$-string is found to be
\eq
\ln(\sin(\pi j/(K+2))/\sin(\pi(j+1)/(K+2)))\,, \qquad j=1,2,\dots,K.
\en
The reason why only $K$ out of the $K{+}2$ anti-Stokes
directions allow for zeros of the wavefunction is that on the two
anti-Stokes lines next to the sector where the initial subdominant
solution is defined (the anti-Stokes lines bordering the union of
segments {\bf 1} and {\bf 2} in the example) there is only a single
`pure' WKB solution, and thus there is no chance to get the
cancellation between two WKB solutions which leads to the wavefunction
zeros near the
other anti-Stokes directions.

It would be interesting to reproduce these results from a study of the
relevant nonlinear integral equations in the integrable models, but we
will leave a more detailed study of such issues for future work.

\resection{The numerical algorithm and the dual formulation of the
boundary problem}
\label{appb}
In this appendix we describe the numerical methods used to test our
conjectures. We solved the pseudo-differential equations using an
iterative power-series method, which we describe below. On the
integrable model side, there is a well-known method to rewrite an
infinite set of Bethe ansatz equations such as (\ref{dall0}) into a
finite set of nonlinear integral equations
(NLIEs)~\cite{Zinn-Justin:1997at}. We applied this to the $K{=}1$
simply-laced cases $A_{n-1}$ and $D_n$.

\subsection{The $\chi$-functions and the generalised Cheng algorithm.}

The special solution
$\psi(x,E,{\bf g})$ and its derivatives can be written in terms of the
$\chi$-functions (\ref{chib}) and the spectral determinants as

\eq
{ d^m \over dx^m}\psi(x,E,{\bf g}) = Q^{(1)}_{[0]}(E, {\bf g}) { d^m
\over dx^m}\chi_0 (x, E, {\bf g})+ Q^{(1)}_{[1]}(E, {\bf g}) { d^m
\over dx^m} \chi_1 (x, E, {\bf g})+\dots~.
\label{psichiq1}
\en
By solving for $Q^{(1)}_{[0]}(E, {\bf g})$:
\eq
Q^{(1)}_{[0]}(E, {\bf g}) = {W[\psi,\chi_1,\chi_2,\dots] \over
W[\chi_0,\chi_1,\chi_2,\dots]}~.
\label{QW}
\en
{}In order to calculate the spectral determinant from~(\ref{QW}) we
must first find the
$\chi$-solutions by numerically solving the pseudo-differential equations.
This can be done, with very high precision, using a generalisation
of the iteration method introduced by Cheng many years
ago~\cite{cheng:1962} and more recently used in~\cite{Dorey:2001uw}.
In the following we shall define the iterative solution for each Lie
algebra on a
case-by-case basis.\\

{\bf $A_{n-1}$}: We begin by defining a solution
for~(\ref{sun0})
\eq
D_n({\bf g}) \chi(x,E,{\bf g})= (-1)^n P_K(x,E)  \chi(x,E,{\bf g})~,
\label{sunn}
\en
and its adjoint equation
\eq
D_n({\bf g^\dagger})\chi^{\dagger}(x,E,{\bf g}^{\dagger})=
P_K(x,E)\chi^{\dagger}(x,E,{\bf g}^{\dagger})~,
\label{sunnq}
\en
where ${\bf g} {=}\{g_{n-1}, \dots,g_1, g_0 \}$ and ${\bf
g^{\dagger}} {=}\{ n{-}1{-}g_0, n{-}1{-}g_1, \dots, n{-}1
{-}g_{n-1} \}$.

Equation (\ref{sunnq}) was defined, as usual, by applying the rule
$(d^p/dx^p)^{\dagger}=(-1)^p(d^p/dx^p)$.
Notice also  that (\ref{sunnq}) is not the
$\psi^{(n-1)}$-related ODE, which is    instead obtained by  simply
replacing ${\bf
g}$ with ${\bf g}^{\dagger}$ in (\ref{sunn}).

Consider equation~(\ref{sunn}). The first step is to define a
linear operator
\eq
L^A_{\bf g}(x^p) = \frac{x^{p+n}}{\prod_{b=0}^{n-1}
  (p+n-g_b )
    }
\en
such that for any polynomial ${\cal P}(x)$ of  $x$
\eq
D_n({\bf g}) L^A_{\bf g}({\cal P}(x))= {\cal P}(x)~.
\label{dna}
\en
Now it is easy to check that
\eq
\chi_a(x,E,{\bf g}) = x^{g_a} +(-1)^n L^A_{{\bf g}} \left( P_K(x,E)
\chi_a (x,E,{\bf g}) \right)
\label{chia}
\en
defines  a solution to  (\ref{sunn}) solvable by iteration. Setting
$g_a^{\dagger}=n-1-g_{n-1-a}$,
\eq
\chi^{\dagger}_a(x,E,{\bf g}^{\dagger}) = x^{g_a^{\dagger}} +
L^A_{{\bf g}^{\dagger}}   \left( P_K(x,E) \chi_a^{\dagger} (x,E,{\bf
g}^{\dagger}) \right)~,
\label{chiab}
\en
instead defines an iterative  solution to the adjoint equation
 (\ref{sunnq}). \\

{\bf $D_{n}$}: The appropriate linear operator for the self-adjoint equation
(\ref{so2n0})
\eq
D_{n}({\bf g^{\dagger}}) \left( \frac{d}{dx} \right)^{-1} D_{n}({\bf
g})  \chi(x,E,{\bf g})=\sqrt{P_K (x,E)} \left(\frac{d}{dx} \right)
\sqrt{P_K (x,E)} \,\chi(x,E,{\bf g})
\label{so2n02}
\en\\
is
\eq
L^D_{\bf g}(x^p) =
\frac{(p+\frac{h^\vee}{2}+1)\,x^{p+h^\vee+1}}{\prod_{b=0}^{2n-1}
  (p+1+\lambda_b )  }~,
\en
and the corresponding  iterative solution is
\eq
\chi_a(x,E,{\bf g} )= x^{\lambda_a} + L^D_{\bf g}\Bigl
(\sqrt{P_K(x,E})\lf(\frac{d}{dx}\ri) \sqrt{P_K(x,E)} \,\chi_a
(x,E,{\bf g} )\Bigr)~.
\label{recdn}
\en
The relation between the $\lambda$'s and the $g$'s is given in
(\ref{lambdag}). \\

$B_{n}$: The iterative solution of the $B_n$ equation~(\ref{so2n10})
\eq
 D_{n}({\bf g^{\dagger}}) D_{n}({\bf g}) \chi(x,E, {\bf g}) =- \sqrt{P_K (x,E)}
\left(\frac{d}{dx} \right) \sqrt{P_K (x,E)} \chi(x,E, {\bf g})
\label{so2n102}
\en
and its adjoint
\eq
 D_{n}({\bf g^{\dagger}}) D_{n}({\bf g})  \chi^{\dagger}(x,E, {\bf g}) = \sqrt{P_K (x,E)}
\left(\frac{d}{dx} \right) \sqrt{P_K (x,E)} \chi^{\dagger}(x,E, {\bf
g})
\label{so2n102a}
\en
are, respectively,
\eq
\chi_a(x,E,{\bf g}) = x^{\lambda_a} - L^B_{\bf g}\Bigl
(\sqrt{P_K(x,E})\lf(\frac{d}{dx}\ri) \sqrt{P_K(x,E)} \,\chi_a
(x,E,{\bf g} )\Bigr)~
\en
and
\eq
\chi_a^{\dagger}(x,E,{\bf g}) = x^{\lambda_a} + L^B_{{\bf g}}\Bigl
(\sqrt{P_K(x,E})\lf(\frac{d}{dx}\ri) \sqrt{P_K(x,E)}
\,\chi_a^{\dagger} (x,E,{\bf g} )\Bigr)~,
\label{bnc}
\en
with
\eq
L^B_{\bf g}(x^p) = \frac{x^{p+h^\vee+1}}{\prod_{b=0}^{2n-1}
  (p+1+\lambda_b )  }~.
\en\\
The relation between the $\lambda$'s and the $g$'s is given in
(\ref{lambdag1}). \\

$C_{n}$:~~The $\chi$-solutions to the self-adjoint equation~(\ref{sp2n0})
\eq
 D_{n}({\bf g^{\dagger}}) \left(\frac{d}{dx} \right)D_{n}({\bf g})\,\chi_a(x,E, {\bf g})
= P_{K }(x,E)  { \left(d \over dx \right)^{-1}}P_{K} (x,E)
\,\chi_a(x,E, {\bf g})
\label{sp2n02}
\en\\
satisfy
\eq
 \chi_a(x,E,{\bf g})= x^{\lambda_a} + L^C_{\bf g}\Bigl (P_K(x,E)\lf(\frac{d}{dx}\ri)^{-1}
P_K(x,E) \,\chi_a(x,E,{\bf g} )\Bigr)
\label{reccn}
\en
with
\eq
L^C_{\bf g}(x^p) = \frac{x^{p+2n+1}}{\prod_{b=0}^{2n-1}
  (p+1+\lambda_b )  }~,
\en
and the relation between the $\lambda$'s and the $g$'s  given in
(\ref{lambdag3}).

\subsection{Dual formulation of the boundary problem}

{}From (\ref{QW}), with the numerical estimates for the
$\chi$-functions and their derivatives at large
values of $x$, replacing  $\psi(x,E,{\bf g})$ by its asymptotic
behaviour and by varying $E$  one has in principle access to
$Q^{(1)}_{[0]}(E, {\bf g})$ and in particular  to its zeros.
This process is certainly possible but it is tedious and the CPU time
increases at least quadratically with the order of the equation.
Surprisingly, there is a short-cut that makes  the algorithm
essentially order-independent.

To see this, consider the
$A_{n-1}$ case

\eq
C Q^{(1)}_{[0]}(E, {\bf g}) =
W[\psi,\chi_1,\chi_2,\dots,\chi_{n-1}]~,
\label{QW1}
\en
where
\eq
C=W[\chi_0,\chi_1,\chi_2,\dots,\chi_{n-1} ] =(-1)^{{\bf int}[n/2]}
\prod_{i=0,j=i+1}^{n-1 }(\lambda_i-\lambda_j)~.
\en
Expand (\ref{QW1}) with respect to the first column:
\eq
CQ^{(1)}_{[0]}(E,{\bf g}) = \sum_{p=0}^{n-1}  (-1)^p \;
w_p\;\frac{d^p \psi}{dx^p}~,
\label{QW2}
\en
where, borrowing the short-hand notation of \S\ref{a3d3},
\bea
w_p(E,{\bf g}) &=&[0,1,\dots,p-1,p+1,\dots, n-1]  \nn\\
&=& {\bf det} \left[ \left(\vec{\chi},\frac{d}{dx} \vec{\chi}
,\dots,\frac{d^{p-1} }{dx^{p-1}}\vec{\chi},\frac{d^{p+1}
}{dx^{p+1}}\vec{\chi},\dots,\frac{d^{n-1} }{dx^{n-1}}\vec{\chi}
\right )\right]
\eea
with $\vec{\chi}= ( \chi_1,\chi_2,\dots, \chi_{n-1} )$~.
 At large  $x$ on the positive real axis
\eq
{ d^p \over dx^p}  \psi(x) \sim (-1)^p \; x^{(1{-}n+2p)M/2}
\exp(-x^{M+1}/(M{+}1)) \quad,\quad (M >K/(h^{\vee}{-}K) ) ~.
\label{asyapp}
\en
Therefore, if for all  functions $w_p(E,{\bf g})$
we have
\eq
w_p(e_i,{\bf g})=o\left(  x^{(1-n)M/2}
\exp(x^{M+1}/(M{+}1))\right)~,~~p =0,1,\dots,n-1
\label{general}
\en
for some $E{=}e_i$, then  from  (\ref{QW2})
\eq
Q^{(1)}_{[0]}(e_i,{\bf g})=0~.
\en
This shows that $\{ e_i \} \subset \{ E_i^{(1)} \}$. We shall now
argue  that the set $\{ e_i \}$ is not empty and further  prove that
\eq
\{ e_i \} = \{ E_i^{(1)} \}.
\label{eqv}
\en
To do this consider
$
w_{n-1}(E,{\bf g})=[0,1,3,\dots,n-2]~.
$
By repeated differentiation
and substitution of the
ODE, one can prove  as in~\cite{Dorey:2000ma}  that
\eq
w_{n-1}(E,{\bf g}) = \Big ( (-1)^{{\bf int}[(n-1)/2]}
\prod_{i=1,j=i+1}^{n-1} (g_i-g_j) \Big)
\chi^{\dagger}_{n-1}(x,E,{\bf g}^{\dagger})
\en
i.e. $w_{n-1}$ satisfies the adjoint $A_{n-1}$-equation
(\ref{sunnq}) and thus can be determined using (\ref{chiab}) with
$g_{n-1}^{\dagger}=n-1-g_0$.

At large $x$\,,
\eq
w_{n-1}(x) =  S(E,{\bf g})\left(x^{(1-n)M/2}+ o(x^{(1-n)M/2})
\right) \exp\left(\fract{x^{M+1}}{M{+}1} \right)+\dots
\label{sumq}
\en
where the dots in (\ref{sumq}) indicate terms that grow
exponentially  at most as
\eq
\exp \left(\cos \left(\fract{2\pi}{n} \right) \fract{x^{M+1}}{M{+}1}
\right)~.
\label{sumqe}
\en
Therefore, the  condition (\ref{general}) for $p{=}n{-}1$ is
fulfilled for any value of $E{=}e_i$ such that $S(e_i,{\bf g})=0$.
Since this a single condition imposed on a nontrivial function of
$E$, we expect an infinite  but countable set of solutions.

For the next step, we want to argue that if the condition
(\ref{general}) is satisfied for $p{=}n{-}1$ it is simultaneously
satisfied  for $p{=}0,1,\dots,n-2$.

{}From (\ref{sumq}) we see that at $S(e_i,{\bf g})=0$ also
\eq
{d^p \over dx^p} w_{n-1}(e_i,{\bf g})=o\left( x^{b}
\exp(x^{M+1}/(M{+}1))\right)
\label{general1}
\en
for any finite complex  number $b$. We now claim that all of the
$w_{a}$ can be written as  a linear combination of
$w_{n-1}$ and its derivatives:
\eq
w_{p}(x,E,{\bf g})= \sum_{a=0}^{n-1-p} {K_a^{(p)}\over x^{n-1-p-a}
}{d^a \over dx^a} w_{n-1}(x,E,{\bf g})~,
\label{true}
\en
where the constants $K_a^{(p)}$ depend only on  ${\bf g}$. It is
very simple   to see that (\ref{true}) is true in general. Write the
adjoint $A_{n-1}$ ODE in the `expanded' form
\eq
\frac{d^n}{dx^n}\, w_{n-1}= \left({A_2 \over x^2}
\frac{d^{n-2}}{dx^{n-2}}+ {A_3 \over x^3}
\frac{d^{n-3}}{dx^{n-3}}+\dots + P_K(x,E) \right)\,w_{n-1}~.
\label{fullode}
\en
Differentiating  once
\eq
{d \over dx} w_{n-1}=[0,1,2,\dots,n-3,n-1]=w_{n-2}~,
\en
taking further derivatives and   using (\ref{fullode}) the  general
result is
\eq
w_{p-1}={d \over dx}  w_p-{A_{n-p} \over x^{n-p}} w_{n-1}~.
\label{systemrec}
\en
Solving (\ref{systemrec}) recursively  proves (\ref{true}).

Equation (\ref{true}) and equation (\ref{general1}) together are
equivalent to (\ref{general}). To prove  (\ref{eqv}), substitute the
asymptotic behaviours~(\ref{asyapp}) and
\eq
w_{n-1-i} \sim {d^{i} \over dx^{i}} w_{n-1}(x) \sim S(E,{\bf g})
x^{(1-n+2i )M/2} \exp\left(\fract{x^{M+1}}{M{+}1} \right)
\en
into (\ref{QW2}) to see that
\eq
C Q^{(1)}_{[0]}(E,{\bf g}) =n S(E, {\bf g})~.
\en
The eigenvalues are obtained by solving
$\chi^{\dagger}_{n-1}(x,E,{\bf g}^{\dagger}){=}0$ for fixed $x$,
chosen large enough to capture the asymptotic behaviour of the
solution while also ensuring the power series is reliable. This
condition is different from the precise requirement
(\ref{general1}), but considering (\ref{sumq}) and (\ref{sumqe}) it
is clear that, provided $x$ is very large, the error can be
minimised and it selects the points on the complex $E$-plane where
the function $S(E,{\bf g})$ is approximately zero.

Similarly, for $B_n$ the eigenvalues of the spectral determinant
$Q_{[0]}^{(1)}$ are found by solving (\ref{bnc})
for $\chi^{\dagger}_{2n-1}(x,E,{\bf g})$.  The situation is more
complicated in the presence of the integral operators that appear
in the
$D_n$ and the $C_n$ models,  and we
have not yet fully completed the analysis. However, guided by the
above and the results when the integral operator is absent, the
obvious prescription to compute $\chi_{2n-1}(x,E,{\bf g})$ from
(\ref{recdn}) and $\chi_{2n+1}(x,E,{\bf
  g})$   from  (\ref{reccn})   for $D_n$ and $C_n$ respectively
works  very well.

\subsection{NLIEs}
The Bethe ansatz equations of type  $A_{n-1}$ and $D_{n}$ for $K{=}1$ can be
 rephrased as a set of $r$ nonlinear integral equations   where $r$ is
 the rank of the algebra~\cite{Zinn-Justin:1997at} (see also
\cite{KP,KBP,DDV,Bazhanov:1996dr,Dorey:1999pv,Dorey:2000ma,Suzuki:2000fc,Dunning:2002cu}).
They are compactly written as
\bea
&& \hskip -25pt
 f^{(a)}(\theta)=
-\frac{2 i \pi}{h^\vee} \frac{\gamma_a}{\alpha}- 2 i  b_0 \,
e^{\theta}  \\[3pt]
&&\hskip -10pt {}+\sum_{b=1}^{r} \left( \int_{ {\cal C}_1} d \theta'
\varphi_{ab}(\theta{-}\theta') \ln(1+e^{f^{(b)}(\theta')}) - \int_{
{\cal C}_2} d \theta' \varphi_{ab}(\theta{-}\theta') \ln(1+e^{-
f^{(b)}(\theta')}) \right)~,\nn
\label{nlie}
\eea
with $b_0{=} m_1 \sin \mu $ and
\eq
\varphi_{ab}(\theta)= \int_{-\infty}^\infty {dk \over 2 \pi} \ e^{i
k \theta}\Bigl (\delta_{ab} - \frac{ \sinh(\pi  \mu
k)}{\sinh(\fr{\pi k}{h^\vee} (h^\vee \mu-1) k) \cosh(\fr{\pi
k}{h^\vee} )} C_{ab}^{-1}(k) \Bigr)~.
\en
The integration contours ${\cal C}_1$ and ${\cal C}_2$ run from
$-\infty$ to $\infty$ just below and just above the real axis
respectively. The algebra dependence is encoded in the constant
$b_0$, the `twists'
$\{\gamma_a /  \alpha\}$, and
the deformed Cartan matrix
\eq
C_{ab}(k) =  \left\{ \begin{array}{ll}
     2  & a{=}b  \\[3pt]
    \frac{-1~}{\cosh( \frac{ \pi k}{h^\vee})} &  {<}ab{>}  \end{array} \right.\,.
\en
Above,  ${<}ab{>}$ implies the nodes $a$ and $b$ of the Dynkin
diagram are connected.  The explicit expressions for the inverse
Cartan matrices can be found in \cite{Zinn-Justin:1997at}.

By construction, the spectrum of the ordinary  differential equation
is encoded in the zeros  $\theta_j$  of $1{+}\exp(f^{(a)}(\theta)$
that lie on the real axis of the complex-$\theta$
plane~\cite{Dorey:1998pt}. The eigenvalues of the relevant ODE are
obtained via the relation
\eq
E_j = e^{\te_j /  \mu}~.
\en

The large-$E$ WKB-like behaviour can be obtained from the
large-$\theta$ limit of these equations.  In this limit the
convolution terms in (\ref{nlie}) can be dropped,  and imposing
$f^{(1)}(\te_j){=} i \pi  (2j-1)$ for integer $j$ yields the
semiclassical prediction
\eq
2 \sin(\mu \pi) \,m_a (E_j^{(a)})^{\mu} \sim (2j{+}1- {2 \over
\alpha h^{\vee}} \gamma_a) \pi~~,~~~ j=0,1,2,\dots
\label{semi}
\en
with $\alpha{=}2(\mu-1/h^{\vee})$.

%

%

%
\end{document}